\documentclass[12pt]{article}
\usepackage{graphicx}
\usepackage{amsmath,amssymb}
\usepackage{amsfonts}
\usepackage{array}
\usepackage{lineno}
\usepackage{braket}
\usepackage{color}
\usepackage{cite}
\usepackage{colortbl}

\newsavebox{\anglearrow}
\savebox{\anglearrow}{%
\begin{picture}(10,15)(0,0)
\put(0,15){\line(0,-1){10}}
\put(0,5){\vector(1,0){10}}
\end{picture}}
\newcommand\pubnumber{}
\newcommand\pubdate{\today}

\def\Title#1{\begin{center} {\Large #1 } \end{center}}
\def\Author#1{\begin{center}{ \sc #1} \end{center}}

\newcommand\pubblock{\rightline{\begin{tabular}{l} \pubnumber\\
         \pubdate  \end{tabular}}}
\newenvironment{Abstract}{\begin{quotation}  }{\end{quotation}}

\begin{document}

\begin{titlepage}
\pubblock

\vfill
\Title{Conceptual Design Report of DaRveX:\\
Decay at Rest $\nu_e$ + Lead Cross Section Measurement \\
at J-PARC MLF \\}
\vfill
\Author{F.~Suekane, Y.~Hino\footnote{Current affiliation: Dept. of Physics, Okayama University, JAPAN}, W.~Noguchi, T.~Tokuraku and N.~Kadota\\ 
 {\small (RCNS ,Tohoku Univ. JAPAN)}, \\
 ~\\
T.~Konno, T.~Kawasaki, Y.~Hoshino and M.~Watanabe \\ 
{\small (Dept. of Physics, Kitasato Univ. JAPAN)},  \\
~\\
Y.~Sugaya \\
{\small (RCNP, Osaka Univ. JAPAN)}, \\
~\\
M.K.~Cheoun\footnote{Not DaRveX member. Contributed section-11.2 \&11.3.} \\
{\small (Dept. of Phys. Soongsil Univ. KOREA)}}
\vfill
\begin{Abstract}
\begin{center}
{\large Abstract} 
\end{center}

DaRveX stands for ``\underline{D}ecay \underline{a}t \underline{R}est \underline{$\nu_e$}-Pb cross (\underline{X}) section measurement,,.
This report explains a conceptual design of DaRveX, an experiment that measures differential cross section of  low energy $\nu_e$+Pb  charged current interaction, 
$$\nu_e (E\sim 30{\rm ~MeV})+ {\rm Pb} \to e^- + xn + {\rm Bi},~~~(x=1{\rm ~or~}2), $$ 
for the first time using $\nu_e$ from $\mu^+$ decay at rest at J-PARC MLF. 

So far, there has not been good $\nu_e$ target to detect low energy $\nu_e$.  
Lead is expected to be an excellent $\nu_e$ target because the cross section is expected to be very large and the delayed coincidence  technique can be used using final state neutron(s) to reduce backgrounds.
However, the cross section have not been measured yet. 
If it is measured, it opens a new window to the future neutrino research field, such as low energy $\nu_e$ oscillation measurements,  flavor specific detection of the supernova explosion $\nu_e$ and understanding of  $\nu_e$-nucleus interactions. 

The DaRveX $\nu_e$ detector  is a remodeling of existing reactor neutrino detector PANDA (Plastic scintillator AntiNeutrino Detector Array).  
A thin lead sheet is sandwiched between two double-layer thin plastic scintillator strips. 
This thin lead-scintillators unit is sandwiched between PANDA scintillator modules and $e^-$ energy and direction will be measured.  
Backgrounds can be reduced by making use of very narrow pulsed beam of MLF, delayed coincidence with neutron absorption signal on Gd,  and requirement of event topology of the charged current $\nu_e$+Pb reaction, together with thick neutron and $\gamma$-ray shields. 
With two years of data taking, the cross section is expected to be measured with 20\% of precision. 

 We performed on-site background measurements at J-PARC MLF in 2021 to understand the real backgrounds. 
 This background measurements are also described in the appendix. 

\end{Abstract}

\end{titlepage}
\def\thefootnote{\fnsymbol{footnote}}
\setcounter{footnote}{0}
\clearpage
\tableofcontents
\clearpage

\section{Introduction}
\label{sec:Intro}

There is an excellent reaction to detect low energy $\overline{\nu}_e$, that is the Inverse Beta Decay (IBD) reaction: 
\begin{equation}
 \overline{\nu}_e + p \to e^+ + n~.
 \label{eq:intro:IBD_reaction}
\end{equation}
Neutrino was discovered by making use of IBD, detecting $\overline{\nu}_e $ from a nuclear reactor~\cite{Cowan56}. 
The reactor neutrino oscillation was discovered\cite{KamLAND02} and three out of the six neutrino oscillation parameters; 
$\theta_{12}$, $\theta_{13}$, $\Delta m_{12}^2$, have been most precisely  measured using IBD. 
All the observed neutrino signals from supernova SN1987A are thought to be produced by IBD~\cite{1987A}.  
Geo-neutrino was firstly observed by IBD~\cite{KamLAND05}. 

The IBD reaction (\ref{eq:intro:IBD_reaction}) is excellent for low energy $\overline{\nu}_e$ detection because, 
(i) the cross section is large, 
(ii) the cross section is directly related to the neutron lifetime and precisely known,  
(iii) the target proton is  abundant in water or organic scintillators,  
(iv) the original neutrino energy can be measured, 
(v) it is a flavor specific reaction, 
(vi) the  final state neutron can be used to remove backgrounds by delayed coincidence technique. 

However, we do not have good $\nu_e$ target to detect low energy $\nu_e$ yet. 
So far low energy $\nu_e$ is detected by the following reaction\cite{LSND0111,KARMEN9800}.
\begin{equation}
 \begin{split}
  \nu_e + {\rm ^{12}C} \to e^- +& {\rm ^{12}N_{gs}}({\rm \tau=11~ms,~ Q=17~MeV}) \\ 
  &~~~~~\searrow  e^+ +{\rm ^{12}C} 
  \end{split}
\end{equation}
However, the reaction cross section is ten times smaller than IBD and the number of carbon  atom is 1/2 of proton in a typical organic liquid scintillator.  
 
If we establish a good low energy $\nu_e$ detection technique, a new window for neutrino research to use the low energy $\nu_e$ will open. 
Lead is expected to be a very good target material to detect low energy $\nu_e$ since it is expected to have very large cross section of the charged current reaction,   
\begin{equation}
 \nu_e + {\rm Pb} \to e^- + xn + {\rm Bi},~~~(x=1{\rm ~or~}2). 
 \label{eq:intro:nue+Pb}
\end{equation}
The neutrons produced can be used for the delayed coincidence to suppress accidental backgrounds. 
However, this reaction has not been used in experiments because the cross section has not been measured yet. 

We are planning to measure differential cross section of the reaction (\ref{eq:intro:nue+Pb}) at J-PARC MLF using the $\nu_e$ produced in the $\mu^+$ decay at rest. 
\begin{equation}
 \begin{split}
 \pi^+({\rm stop,~\tau=26 ns}) \to&  \mu^+ + \nu_\mu \\
 &~~
\searrow
 \mu^+ ({\rm stop,~\tau=2.2\mu s}) \to  e^+ + \underline{~\nu_e~} + \overline{\nu}_\mu
  \end{split}
  \label{eq:Intro:pi+Decay_Chain}
\end{equation}
This experiment will detect the electron produced in the reaction (\ref{eq:intro:nue+Pb}) and measures its energy  and direction\footnote{COHERENT group is measuring the inclusive cross section of $\nu_e$+PB $\to n$+ X at SNS using 1~ton 
lead target~\cite{Daughhetee20} but they will detect only the final state neutron and $e^-$ signal will not be seen.}. 

\section{Possible Physics Studies}
\label{sec:PossiblePhysics}

\begin{figure}[htbp]
\centering
\includegraphics[width=120mm]{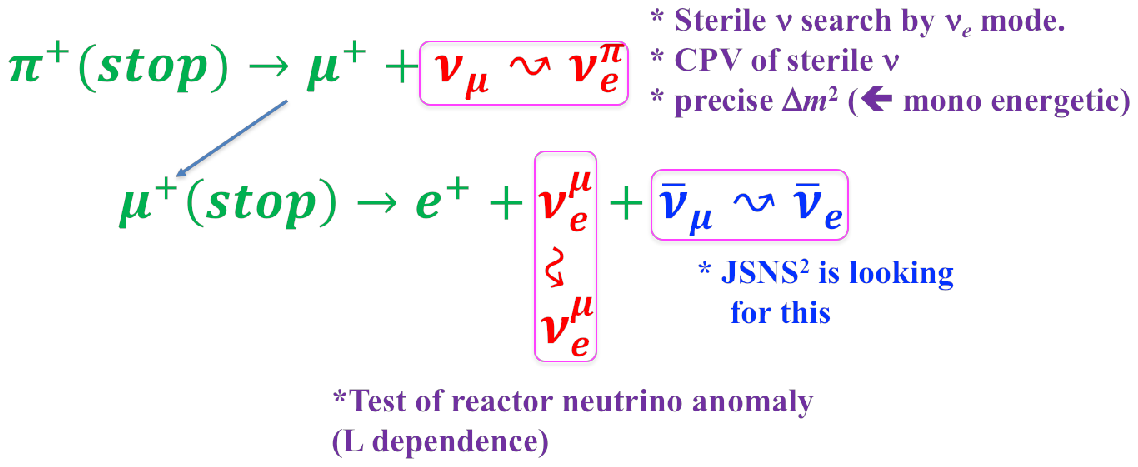}
\caption{\small{
Possible oscillations of decay at rest neutrinos\cite{Suekane2002}.
We call $\nu_e$ in $\mu^+$ decay as $\nu_e^\mu$ and $\nu_e$ generated by oscillation of $\nu_\mu$ in the $\pi^+$ decay as $\nu_e^\pi$ to distinguish them in the discussions.
JSNS$^2$ experiment is looking for sterile neutrino observing $\overline{\nu}_\mu \to \overline{\nu}_e$ appearance~\cite{JSNS2-TDR17}.
}}\
\label{fig:intro:DaR_nu_Oscillation}
\end{figure}

\begin{figure}[htbp]
\centering
\includegraphics[width=14cm]{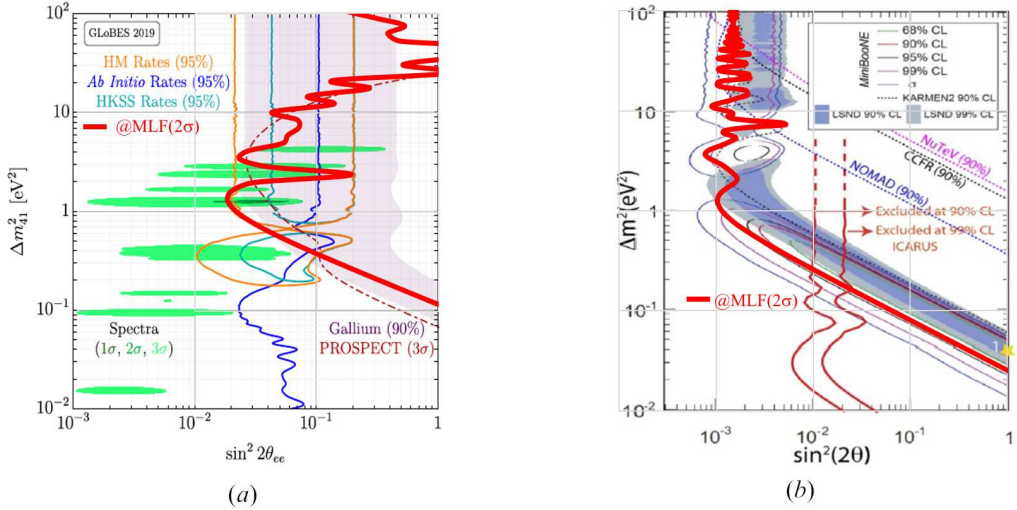}
\caption{\small{Thick red lines show statistical $2\sigma$-significant $\sin^22\theta$ upper limit in possible future DAR $\nu_e$ experiments. 
$(a)$ is for the $\nu_e^\mu$ disappearance measurement and $(b)$ is for the $\nu_\mu^\pi \to \nu_e^\pi$ appearance measurement. 
For both cases, a far detector with 100~ton lead target locates at 30~m baseline and a near detector with 11~ton lead target at 10~m baseline, are assumed as the experimental configuration. 
1~MW beam operation and 5 years of data taking period are also assumed. 
See Appendix-2 for more details. 
}}
\label{fig:MainPart:nue_app_disapp_sensitivity}
\end{figure}

An example of possible  physics study with low energy $\nu_e$ is measurement of low energy $\nu_e$ oscillations. 
$\nu_\mu \to \nu_e$ appearance and $\nu_e \to \nu_e$ disappearance oscillations may be occurring  for decay at rest neutrinos at a baseline $\sim$ 10~m as shown in 
Fig.~{\rm \ref{fig:intro:DaR_nu_Oscillation}} if sterile neutrino exists.
The $\nu_e$ disappearance is the CPT inverted process of $\overline{\nu}_e$ disappearance and $\nu_e^\mu$ can be used to test the reactor neutrino anomaly \cite{Dentler17}. 
Since the energy of the decay at rest $\nu_e$ is ten times larger than that of reactor $\overline{\nu}_e$, it is possible to measure baseline dependence of the oscillation for large $\Delta m^2$ region, where it is difficult to measure by reactor $\overline{\nu}_e$. 
It is also important to measure $\nu_e$ appearance since positive $\nu_e$ appearances at $\Delta m^2 \sim 1{\rm eV^2}$ are reported \cite{MiniBooNE21,LSND98} as well as $\overline{\nu}_e$ appearance. 
The $\nu_\mu \to \nu_e$ oscillation is the CP inverted process of $\overline{\nu}_\mu \to \overline{\nu}_e$ oscillation. 
If CP symmetry is violated, the $\nu_e$ appearance probability and $\overline{\nu}_e$ appearance probability can be different and $\nu_e$ appearance measurement has its own significance. 
CP violation can be measured by comparing the $\nu_e$ appearance probability and $\overline{\nu}_e$ appearance probability.  
Therefore, if $\overline{\nu}_e$ appearance is confirmed at $\Delta m^2 \sim 1{\rm eV^2}$, the $\nu_e^\pi$ appearance measurement will become one of the most important subjects in elementary particle physics. 
Figs.~\ref{fig:MainPart:nue_app_disapp_sensitivity} show statistical sensitivities of possible future $\nu_e$ oscillation measurements. 
These measurements of CP violation is largely systematic free as described in Appendix-2.

The energy range of the decay at rest $\nu_e$ is similar to that of $\nu_e$ generated in supernova explosions. 
Therefore, once the $\nu_e + {\rm Pb}$ cross section is measured, Pb will become a useful target material for the supernova $\nu_e$ detector. 
There actually exists supernova $\nu_e$ detector that uses 79~ton of lead target, called HALO\cite{HALO15}. 
The cross section to be measured here will also be beneficial to the experiment. 
$\overline{\nu}_e$+Pb reaction is expected to be suppressed much due to the Pauli blocking\cite{Scholberg12} and neutral current interaction is also expected to be suppressed\cite{Engel18}. 
The angular distribution of the $e^-$ may be backward peak as described in Section-\ref{sec:DAR_Property}. 
If those properties are confirmed, it becomes possible to  detect flavor specific, directional measurement of the supernova explosion $\nu_e${\rm \cite{SNS05,HALO15}}. 
The  directional measurement of neutrino is especially important because the neutrino is expected to be released from supernova  explosion before the light  and the $\nu_e$ measurement can inform astronomers where to look before the light arrives (SNEWS)\cite{SNEWS20}.  
It is difficult to measure the direction of low energy $\overline{\nu}_e$ since the angular distribution of the final state $e^+$ in IBD reaction (\ref{eq:intro:IBD_reaction}) is almost isotropic. 
Therefore, it is an important subject to measure the angular distribution of $e^-$ produced in the reaction (\ref{eq:intro:nue+Pb}). 

Cross sections of neutrino-nucleus interactions at this energy range have rarely been measured so far and the differential cross section measured here shall give precious information to nuclear physics{\rm \cite{SNS05}}, such as nuclear responses to the weak interactions, neutron skin thickness, nucleosynthesis in supernova explosion for some p-nuclei, etc. 
See Appendix-2 for more details. 
In addition, once the experimental technique of this $\nu_e$-Pb cross section measurement is established here, cross section measurements of other  nuclei may become possible by replacing the lead sheets by the sheets made of other materials.  

\section{$\nu_e$-nucleus cross section}
\begin{figure}[htbp]
\centering
\includegraphics[width=90mm] {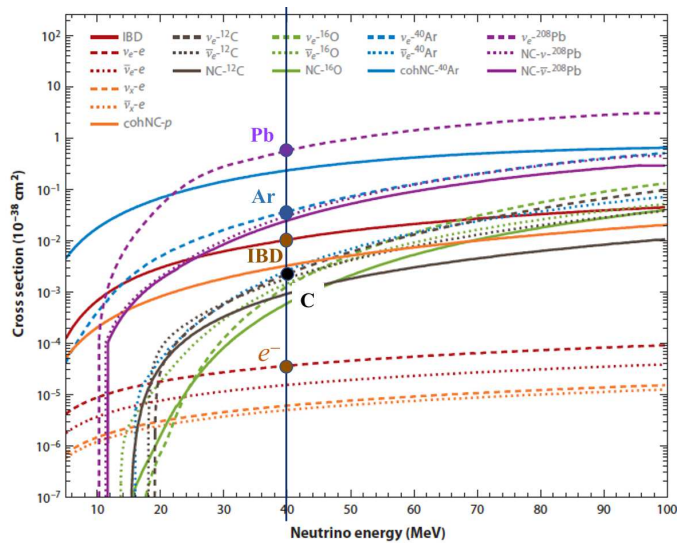}
\caption{\small{Various $\nu$-A cross sections calculated~\cite{Scholberg12}.
The reference energy 40~MeV is a  typical energy of $\nu_e$ produced in $\mu^+$ decay at rest. 
}}
\label{fig:nu-A_Xsection}
\end{figure}

So far low energy $\nu_e$ has been measured by LSND and CARMEN experiments using $\nu_e$-${\rm ^{12}C}$ interactions in their liquid scintillators.
However, $\nu_e$-${\rm ^{12}C}$ cross section is 1/8 of  the inverse beta decay (IBD) cross section for 40~MeV neutrinos as shown in Fig.~\ref{fig:nu-A_Xsection}.
Moreover, the number of $^{12}$C nucleus is 60\% of that of the proton in  a typical organic (${\rm \sim (CH_{1.6})_n}$) liquid scintillator and the event rate of $\nu_e$-${\rm ^{12}C}$ interaction is 1/15 of  the IBD interaction for the same $\nu_e$ and $\overline{\nu}_e$ fluxes. 
On the other hand, as Fig.~\ref{fig:nu-A_Xsection}  shows, lead is expected to have very large cross section. 
Table~\ref{tab:Intro:nu-A_EventRate} compares the cross sections and event rates of various target materials. 
\begin{table}[htbp]
\small{
\begin{center}
\begin{tabular}{|c||c|c|c|c|c|c|}  
\hline
         & $\sigma$@40~MeV          &           & $\rho$         & $A_{\rm eff}$ & $N_\nu$@10m         & Ratio   \\
 Target        & [${\rm 10^{-40}cm^2}$]  &  Target form  & [g/cm$^3$]  &         -     & [/ton/MW/day] &  (target mass)        \\
 \hline
 \hline
  p (IBD) &  1.6 &liquid scinti. &  0.86 &  8.5 & 33 & 0.68 \\ 
\hline  
\hline  
${\rm ^{208}Pb}$  & 57 & raw material & 11.3& 207 & 48 & 1 \\
\hline
 ${\rm ^{12}C}$  & 0.2 & liquid scinti. & 0.86& 13.6 & 2.5 &0.053   \\
\hline
 ${\rm ^{40}Ar}$  & 3 &liquid Ar & 1.39 & 40 & 20 &0.27  \\
\hline
 ${\rm ^{56}Fe}$  & 4.7\cite{Gaponov04} & raw material & 7.87 & 55.8  & 10 & 0.21   \\
\hline
 $e^-$  & 0.0038 & ${\rm H_2 O}$ & 1.0 & 2.25 & 0.29 &0.0061   \\
\hline
 \end{tabular}
\caption{\small{
Comparison of event rate of various target materials. 
$\sigma$ is the cross section for 40~MeV $\nu_e$ (for IBD case, $\overline{\nu}_e$).
$\rho$ is the target density, where liquid scintillator (LAB  ${\rm (CH_{1.6})_n}$) is assumed as target material 
for proton (IBD) and ${\rm ^{12}C}$.
$A_{\rm eff}$ is the effective atomic weight  which is defined as $N_{\rm A}/A_{\rm eff}$ being the number of target nuclei in 1[g] target, where $N_{\rm A}$ is the Avogadro number. 
$N_\nu$ is the event rate per 1[ton] target assuming neutrino flux is $f_\nu^* = 4.4\times 10^{12}[{\rm /cm^2 /day}]$ 
(see Eq.~(\ref{eq:reffnu}):  the expected flux at 10[m] from the mercury target with 1[MW] operation of MLF).  
Ratio(target mass) is the relative event rate per target mass.   
Small $\nu_e$-$e^-$ (in ${\rm H_2O}$ target) elastic scattering is also shown as a comparison. 
}}
\label{tab:Intro:nu-A_EventRate}
\end{center}
}
\end{table}
The event rate for lead target per unit weight is 19 times larger and per unit volume is 240 times larger than ${\rm ^{12}C}$ target in liquid scintillator. 

Due to those properties, there have been a lot of interest for low energy $\nu_e$-Pb interactions regarding the studies on, not only the neutrino physics but also supernova neutrino detection and neutrino-nucleus interactions
 \cite{
 Pattavina20, 
 Ejiri19, 
 Albert18, 
 Hedayatipoor18, 
 Engel18, 
 Civitarese16, 
 Almosly16, 
HALO15, 
 Scholberg12, 
 Vaananen11, 
 SNS05, 
 Reitzner05, 
 Kolbe03, 
 Boyd03, 
 Suzuki03, 
 Jachowicz02,  
 Bacrania02, 
 Zach02, 
 Jachowicz02-2, 
 Volpe02, 
 Kolbe01, 
 Smith01, 
 Elliott00, 
 Elliott00-2, 
 Doe00, 
 Fuller99, 
 Smith97, 
 Hargrove96, 
 Cline94}. 
 There is a number of cross-section calculations \cite{Albert18, Engel18, Scholberg12} and an actual supernova neutrino detector utilizing 79 tons lead target has already been constructed\cite{HALO15}.

\section{Properties  of DAR $\nu_e$ and  $e^-$ produced  }
\label{sec:DAR_Property}
\begin{figure}[htbp]
\centering
\includegraphics[width=80mm]{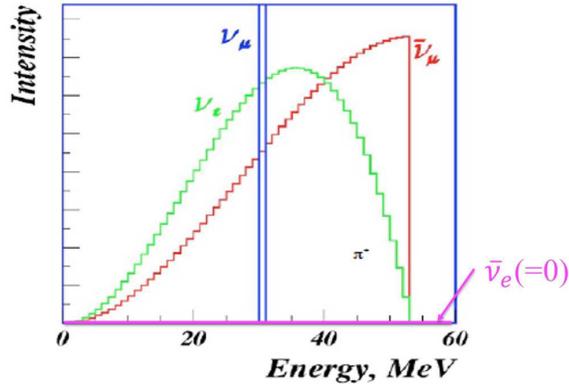}
\caption{\small{
Energy spectra of neutrinos produced $\pi^+$ and $\mu^+$ decay at rest\cite{Bolozdynya1211}.
It is also important that there is no $\overline{\nu}_e$ in the decay chain to perform $\overline{\nu}_e$ appearance experiment. 
}}
\label{fig:intro:DaR_nu_spectra}
\end{figure}

In the decay chain of $\pi^+ \to \mu^+$ at rest,  $\nu_\mu$, $\overline{\nu}_\mu$ and $\nu_e$ are produced as shown in Eq.~(\ref{eq:Intro:pi+Decay_Chain}).
The energy spectra of each neutrino is known precisely as shown in Fig.~\ref{fig:intro:DaR_nu_spectra}. 
$\nu_e$ is produced in the decay of $\mu^+$ at rest. 
The average energy of $\nu_e$ is $\sim$ 32~MeV. 
We use this $\nu_e$ in this experiment. 
The total flux of each neutrino is the same. 
It means if $\nu_e$ flux is measured by lead target, the fluxes of $\nu_\mu$ and $\overline{\nu}_\mu$, which are generally difficult to measure by other methods, can be known.

\begin{figure}[htbp]
\centering
\includegraphics[width=90mm]{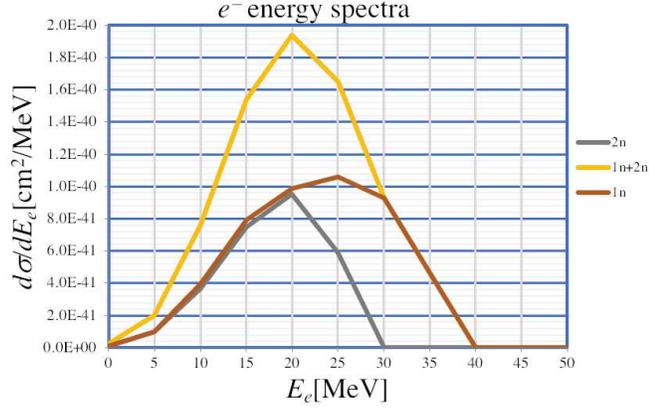}
\caption{\small{
An approximate energy spectra of $e^-$ generated in the reaction (\ref{eq:intro:nue+Pb}) with incident $\nu_e^\mu$.
The brown line is for one neutron emission reaction, the gray line is for two neutrons emission and yellow line is for the sum of both. 
According to \cite{Engel18}, one neutron emission probability is  63\% and two neutron emission probability is 37\%.
}}
\label{fig:Intro:e-_energy_spectra}
\end{figure}

Fig.~\ref{fig:Intro:e-_energy_spectra} shows expected energy spectra of the electron  emitted in the reaction (\ref{eq:intro:nue+Pb}) with  incident $\nu_e^\mu$, calculated in  App.-2 (Fig.~\ref{fig:APP1:nue_e-_spectra}$(b)$). 
The average energy is 20~MeV. 
The energy threshold of the electron signal will be set at $\sim$20~MeV to remove $\gamma$-ray backgrounds from nuclei which absorbed neutron. 
This threshold energy is also much higher than the energy of the background $\gamma$s from the natural radio isotopes, such as $^{238}$U, $^{232}$Th and $^{40}$K. 

\begin{figure}[htbp]
\centering
\includegraphics[width=80mm]{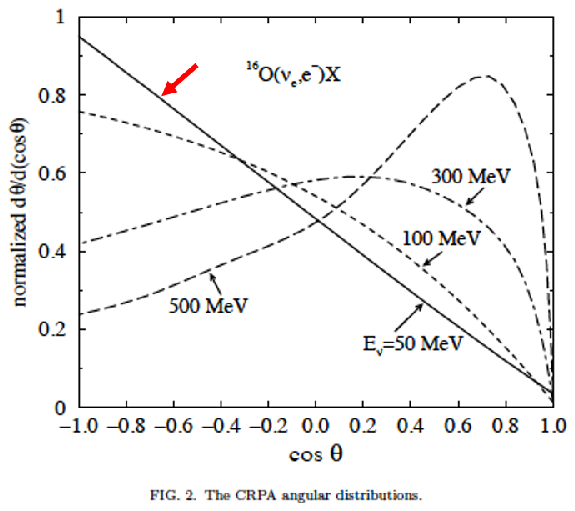}
\caption{\small{Angular distribution of $\nu_e + {\rm ^{16}O} \to e^- +X $ reactions for various $\nu_e$ energies \cite{Kolbe03}. 
The red arrow shows that the $e^-$ emission is predominantly in the backward direction for low energy (50~MeV) neutrino.  
}}
\label{fig:MainPart:e-_angle}
\end{figure}

If the angular distribution of $e^-$ in the reaction (\ref{eq:intro:nue+Pb})  is not isotropic,  neutrino direction can be measured from the $e^-$ emission angle.  
This property is very important for supernova explosion neutrino detection since the direction of the supernova can be known before the lights arrive. 
If the elementary process of the reaction (\ref{eq:intro:nue+Pb}) is the interaction between $\nu_e$ and a  constituent quark with a mass of hundreds MeV,  a backward peak of the $e^-$ emission is expected. 
Fig.~\ref{fig:MainPart:e-_angle} shows a calculation of the angular distributions of electron in the charged current reaction ${\rm ^{16}O} + \nu_e \to e^- + X$\cite{Kolbe03}. 
It shows that for decay at rest neutrino energy, the emission angle of $e^-$ is predominantly backward.
On the other hand, if the reaction (\ref{eq:intro:nue+Pb}) proceeds by producing excited Bi and then the neutron is emitted from the decay of the excited Bi, the electron emission becomes isotropic.
In any case, it is important to measure the direction of the electron emission and check if there is any anisotropicity of the angular distribution for future application of this technique. 
 
 We have seen that the $\nu_e$+Pb interaction is expected to have very good properties to detect low energy 
$\nu_e$. 
However, the properties discussed above are based on theoretical calculations which suffer from ambiguities of nuclear effects.  
It is necessary to experimentally measure $\nu_e$+Pb interaction quantitatively to make use of such good properties for future experiments and we are designing the following experiments. 

\section{$\nu_e$ from J-PARC MLF}
\label{sec:nue_from_MLF}
\begin{figure}[htbp]
\centering
\includegraphics[width=14cm]
{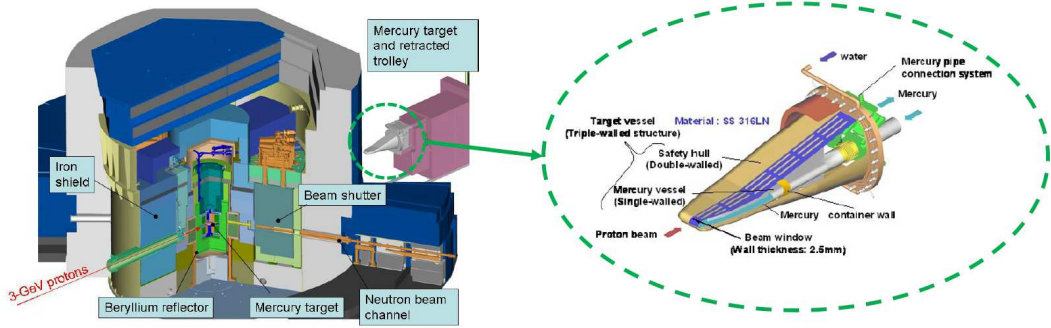}
\caption{\small{
A schematic drawing of the J-PARC spallation neutron source (left) and the mercury target (right) \cite{JSNS2-TDR17}. 
}}
\label{fig:Intro:Hg_target}
\end{figure}

Figs.~\ref{fig:Intro:Hg_target} show a schematic drawing of the J-PARC spallation neutron source (left) and large mercury target (right) \cite{JSNS2-TDR17}. 
1~MW (design value),  3~GeV pulsed proton beam comes in from the left and hits the mercury target and produces pions. 
 The $\nu_e$ to be used in this experiment is generated in the $\mu^+$ decay in the decay chain shown in Eq.~(\ref{eq:Intro:pi+Decay_Chain}). 
$\pi^+$ decays after stopping in the mercury target.
 The kinetic energy of $\mu^+$ produced in the $\pi^+$ decay at rest is only 4.1~MeV and it stops within a range of  
 $\sim$0.1~mm in the target. 
 Therefore, most of $\mu^+$ decay after stopping in the target, too.   
 
\begin{figure}[htbp]
\centering
\includegraphics[width=120mm]{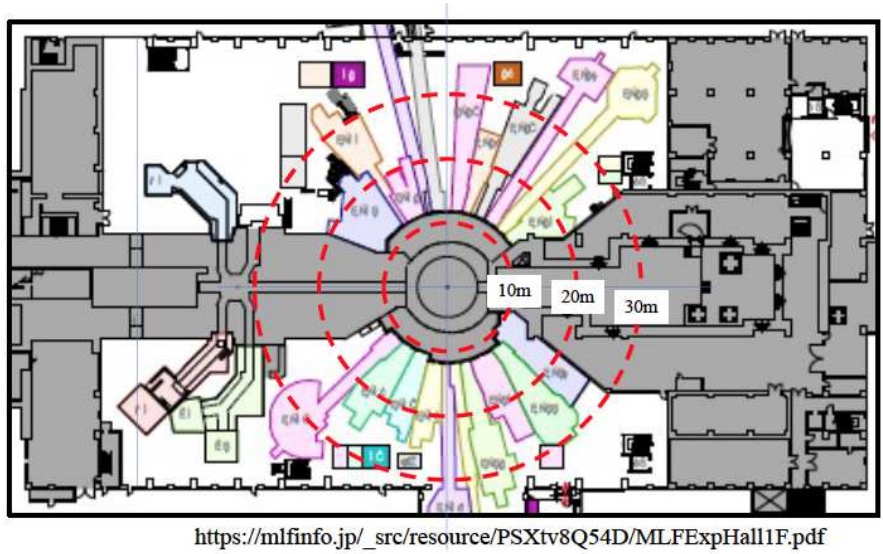}
\caption{\small{
A floor plan of the MLF first floor neutron beamlines.
3~GeV proton beam comes in from the left and hits the mercury target at the center. 
The distance from the mercury target is overlaid by red dotted line circles. 
The area at 10~m distance is an open stage on heavy concrete shield. 
}}
\label{fig:Intro:MLF_1F}
\end{figure}

 Fig.~\ref{fig:Intro:MLF_1F} shows the neutron beamlines of MLF on the first floor and distance from the mercury target.
The size of DaRveX $\nu_e$ detector is typically (1~m)$^3$. 
Since neutrinos can not be stopped by any materials, the detector can be placed anywhere and need not to use existing beamline. 
We are aiming at placing the detector at a distance $L_B\sim 10$~m. 
In this report we assume we can set the detector at a distance 10~m from the Hg target and evaluate performances of the experiment. 

In Appendix.-1, the expected $\nu_e$-Pb reaction rate is calculated  as $N_{\nu_e}^* \sim 47.9[\nu_e /{\rm day}]$ (see Eq.~(\ref {eq:Intro:hat(nnue)}))
 for the reference parameter values shown in Table~\ref{tab:Signal:Reference_Value}, such as 
the beam power: $P_{\rm B}^*$=1~MW, 
the proton energy: $E_{\rm p}^*$=3~GeV, 
the baseline: $L_{\rm B}^*$=10~m, 
and the target lead mass: $M_{\rm Pb}^*$=1~ton, where the asterisk (*) on the parameters mean they are reference values. 
For the actual experiment, we assume $P_B$=0.8~MW, $L_B$=10~m and $M_{\rm Pb}$=250~kg, and the event rate becomes
\begin{equation}
  n_{\nu_e^\mu {\rm Pb}}=47.9[\nu_e /{\rm day}]\left( \frac{0.8{\rm MW}}{1{\rm MW}} \right) \left( \frac{250{\rm kg}}{1000{\rm kg}} \right)
    = 9.6 [{\rm  /day}]~.
    \label{eq:n_nuPb=}
\end{equation}
With 200[day/year]  of data taking period, 1,900 $\nu_e$-Pb reactions are expected to happen per year. 

\begin{figure}[htbp]
\centering
\includegraphics[width=80mm]
{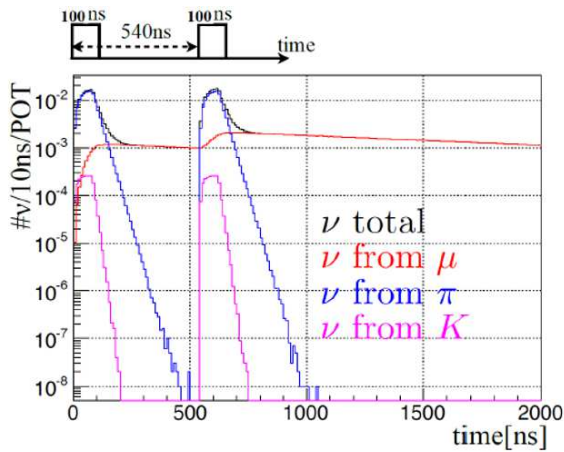}
\caption{\small{
The time structure of the proton beam and that of the neutrinos generated in pion, muon and kaon decays. 
The repetition rate of the twin pulses is 25~Hz \cite{JSNS2-TDR17}. 
The $\nu_e$ is produced as ``$\nu$ from $\mu$,, line.  
By selecting time$>$1.5~$\mu$s, other neutrinos and most of on-bunch backgrounds can be eliminated while keeping $60\%$ of $\nu_e^\mu$. 
}}
\label{fig:Intro:beam_timing}
\end{figure}

\begin{figure}[htbp]
\centering
\includegraphics[width=100mm]
{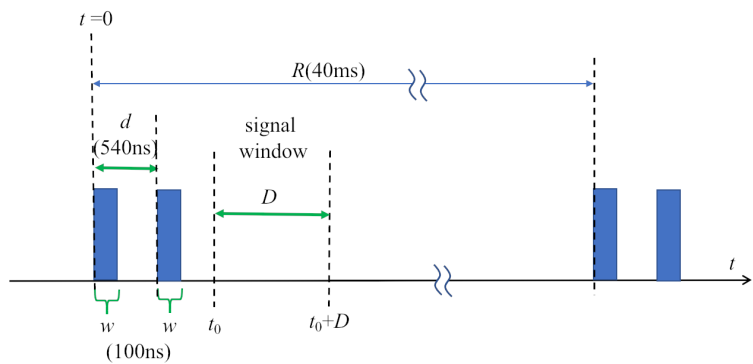}
\caption{\small{
Time structure of the proton beam and the parameters of the timing cut. 
$w=100$ns, $d=540$ns. 
We can optimize $t_0$ and $D$ for $\nu_e$ detection.}}
\label{fig:Intro:Timing_cut}
\end{figure}

The timing structure of MLF beam has extremely good properties for decay at rest neutrino experiments.  
Fig.~\ref{fig:Intro:beam_timing} shows the timing of the beam pulses and the time structure of the neutrino productions. 
A pair of two narrow (100~ns) pulses with separation 540~ns comes every 40~ms. 
By opening the time window, for example, $t_0=1.5 {\rm \mu s}$ and $D=4.0{\rm \mu s}$  in Fig.~\ref{fig:Intro:Timing_cut} to detect $\nu_e$ from $\mu^+$ decay at rest, it is possible to remove the on-bunch beam-associated backgrounds. 
The beam-uncorrelated backgrounds can be reduced to $1\times 10^{-4}$.
The efficiency of $\nu_e$ detection for the time window is, 
\begin{equation}
\epsilon_{bt} \sim \frac{1}{2} e^{-\left((t_0-(w/2))/\tau_\mu \right)}
 \left(1+e^{\left( d/ \tau_\mu \right)} \right)
 \left(1-e^{-\left(D/\tau_\mu \right)} \right)
 = 60.6\%
 \label{eq:Intro:et}
\end{equation}
where $\tau_\mu \sim 2.2\mu s$ is the muon lifetime. 
The beam uncorrelated backgrounds, such as cosmic-ray related background contaminations can be estimated precisely by using off-bunch timing events and can be statistically subtracted from the $\nu_e$ event sample.  
The time dependence of the reduction of the $\nu_e$ event rate corresponds to the muon lifetime can also be used to separate the signal from background statistically. 

The very narrow pulses are especially beneficial to detect neutrinos from $\pi^+$ and $K^+$ decay at rest which may be used in future experiments. 
By setting the  signal time window just the same as the pulses; 100ns$\times$2, the beam-uncorrelated backgrounds can be reduced to $5\times 10^{-6}$ for detecting neutrinos from decays of $\pi$ and $K$ at rest.

\begin{figure}[htbp]
\centering
\includegraphics[width=140mm]
{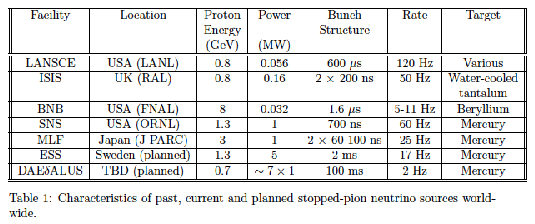}
\caption{\small{
Comparison of various decay at rest neutrino sources in the world. 
\cite{Bolozdynya1211}
}}
\label{fig:Intro:DaR_source}
\end{figure}

 Fig.~\ref{fig:Intro:DaR_source} compares the characteristics of the decay at rest neutrino sources in the world.  
J-PARC MLF has advantages of high power and low duty cycle pulsed beams and heavy target material. 
Therefore, it is ideal to perform DaRveX experiment here.

\section{The DaRveX $\nu_e$ Detector}
\label{sec:DaRveX_Detector}

The DaRveX $\nu_e$ detector will be made by remodeling existing PANDA(Plastic Anti-Neutrino Detector Array) reactor neutrino detector~\cite{PANDA19, PANDA12, PANDA14} and this experiment can be performed very cost effectively. 
PANDA was operated several years and successfully detected reactor $\overline{\nu}_e$ at above ground at a baseline 45~m from an Ohi nuclear reactor. 
It means the various properties of the core part of the DaRveX detector is already understood well. 

\subsection{PANDA reactor $\overline{\nu}_e$ detector}

\begin{figure}[htbp]
\centering
\includegraphics[width=14cm] {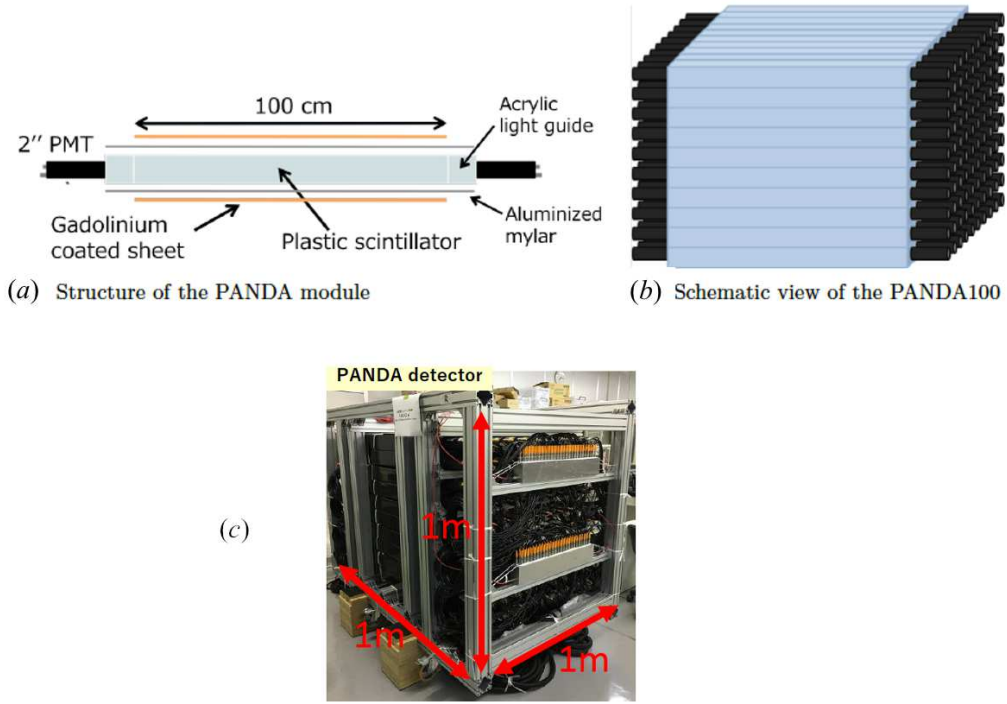}
\caption{\small{$(a)$ A unit of the PANDA plastic scintillator module. 
The size of the scintillator block is ${\rm 10cm\times 10cm \times 1m}$. 
The scintillator block is viewed by 2~inch PMTs from both sides through acrylic light guides. 
The scintillator is wrapped with a paper sheet which contains ${\rm 4.9mg/cm^2}$ of gadolinium(Gd). 
$(b)$ Array structure of the PANDA detector\cite{PANDA19}.  
The PANDA detector consists of 100 scintillator units and the active mass is $\sim$1~ton. 
$(c)$ A picture of the assembled PANDA detector \cite{Konno19}. 
}}
\label{fig:Intro:PANDA}
\end{figure}

Figs.~\ref{fig:Intro:PANDA} show the PANDA detector. 
It consists of 10$\times$10 array of 10cm$\times$10cm$\times$100cm plastic scintillator blocks (BC408\cite{BC408}). 
The reactor neutrino ($\overline{\nu}_e$), whose typical energy is 4~MeV, performs inverse beta decay (IBD) reaction with free proton in the plastic scintillator as shown in Eq.~(\ref{eq:intro:IBD_reaction}). 
Each PANDA plastic scintillator block is wrapped with a paper sheet which contains ${\rm 4.9mg/cm^2}$ of gadolinium(Gd), the corresponding occupancy of the thermal neutron absorption cross section is 0.91/layer. 
The neutron produced in the reaction (\ref{eq:intro:IBD_reaction}) quickly thermalizes colliding with protons in the plastic scintillators and is eventually absorbed by Gd and the excited Gd emits $\gamma$-rays whose total energy is 8~MeV, typically 
30$\mu$s after the IBD reaction. 
By taking the delayed coincidence between the positron signal and Gd signal, PANDA could reduce the backgrounds significantly  and successfully detected reactor neutrinos from Ohi reactor-4\cite{Konno19}.
The PANDA detector is a compact and mobile system.  
While taking neutrino data in Ohi nuclear power plant, the PANDA detector was left alone for several weeks in a truck without any access. 
The DaRveX $\nu_e$ detector inherits such the mobile and maintenance-free properties. 

\subsection{DaRveX conceptual detector structure}
\begin{figure}[htbp]
\centering
\includegraphics[width=10cm] {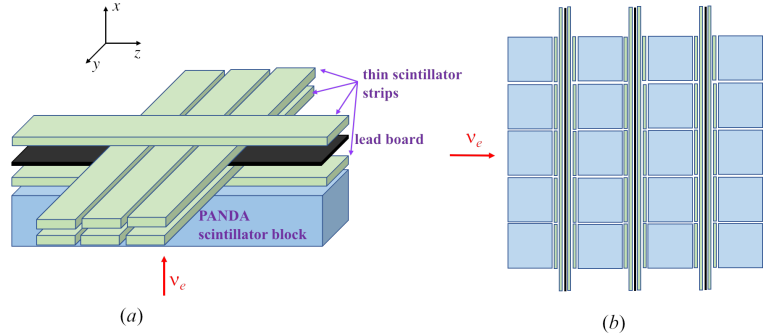}
\caption{\small{ $(a)$  A unit of DaRveX detector.  
A lead board and thin plastic scintillators are sandwiched between PANDA plastic scintillator blocks. 
The arrow labeled $\nu_e$ is the direction of incoming $\nu_e$ beam. 
$(b)$ Cross section of a part of the DaRveX detector. 
 }}
\label{fig:Intro:CDR_scinti_unit}
\end{figure}

Fig.~\ref{fig:Intro:CDR_scinti_unit}$(a)$ shows a unit of the DaRveX detector to detect $\nu_e$-Pb interaction (\ref{eq:intro:nue+Pb}). 
In the detector,  a 4~mm thick lead plate is  sandwiched between two layers of 1cm thick plastic scintillator strips and the 
lead-scintillator strip unit is  inserted between the PANDA scintillator blocks.
We assume to use $8\times 8$ PANDA modules. 
In this case, the lead mass is
\begin{equation}
M_{\rm Pb}={\rm 100cm \times 80cm \times 0.4cm \times 11.34g/cm^3 \times 7layers \sim 250kg}.  
\end{equation}

The PANDA scintillator blocks are read out from both sides by 2 inch PMTs and the scintillator strips are read out by wavelength shifting fibers and MPPC from both sides.
It is possible to measure the hit position along the scintillator and efficient position cuts are possible. 
The detector is surrounded by cosmic ray anti-counters and  shielded by lead bricks and boron sheets heavily.

The energy scale calibration at $\sim 30$~MeV energy range will be performed by the Michel electron and the stability of the gain will be monitored by using the minimum ionizing peak of the penetrating cosmic-ray muons. 

\subsection{An example of $\nu_e$ signal}
Detailed properties of the $\nu_e$ events are discussed in Appendix-1. 
Fig.~\ref{fig:Intro:Principle} shows the  structure of the conceptual detector unit and a typical example of the $\nu_e$+Pb interaction.  
\begin{figure}[htbp]
\centering
\includegraphics[width=140mm] {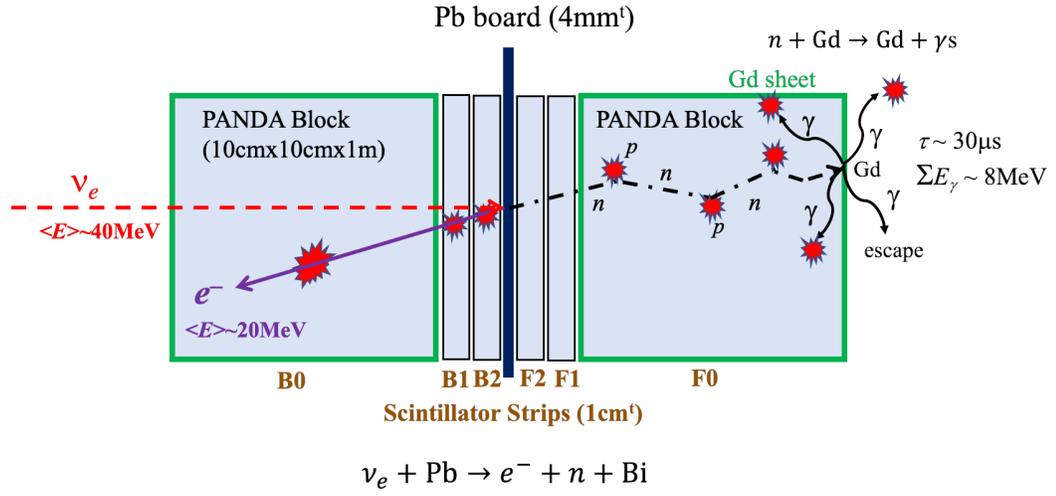}
\caption{\small{
Structure of the conceptual detector unit and  a typical example of $\nu_e$+Pb event. 
$e^-$ ($\braket{E_e}\sim 20$~MeV) generated by the $\nu_e$+Pb interaction produces a triple coincidence of B2, B1 and B0 scintillators, while 
there is no signal in F2 scintillator. 
The  final state neutron  ($\braket{E_n}\sim 4$~MeV) is scattered by protons with mean free path $\sim$10~cm and transfers half of its energy to each proton in average.  
The scattered protons generate small ($\sim$MeV) signal a few~ns after the electron signal. 
The neutron thermalizes after repeating the scattering in the plastic scintillators and eventually be absorbed by Gd contained in the paper sheet. 
The excited Gd emits $\gamma$-rays whose total energy is 8~MeV, typically 30~$\mu$s after the $e^-$ signal.
The $e^-$ signal and the Gd signal form the delayed coincidence.  
The scattered proton signal gives additional information to select $\nu_e$-Pb events.
}}
\label{fig:Intro:Principle}
\end{figure}
The $\nu_e$ hits the Pb target and performs the reaction (\ref{eq:intro:nue+Pb}). 
The average energy of the $\nu_e$ which produces a free neutron in the final state is 40~MeV and 
the average energy of the electron produced is 20~MeV. 
The electron produces triple coincidence of B2, B1 and B0. 
   Since at least 35~MeV is necessary  for $e^-$ to penetrate 14~cm thick plastic scintillators and 4~mm thick lead, the prompt signal is contained within one unit layer of the detector and the event topology is localized. 
 This property helps to remove the fast neutron and accidental backgrounds which tend to show spread signal distributions. 
 
The neutron emitted in the reaction (\ref{eq:intro:nue+Pb}) is scattered by protons many times and eventually thermalizes and 
 absorbed by Gd with mean life time $\sim 30~\mu$s.
 The excited Gd  generates $\gamma$-rays whose total energy is 8~MeV. 
Some $\gamma$-rays from Gd are absorbed by the lead boards and some escape from the detector.  
 Therefore, the visible energy of the Gd signal does not form clear 8~MeV peak. 
 
 According to the reference\cite{Engel18},  37~\% of the events produce 2 neutrons in the final state.  
 For such events, it is possible to reduce the background strongly by requiring double delayed coincidences.

\subsection{$\nu_e$ event selection and efficiency}
\label{sec:main:event_selection}
The $\nu_e$ event is identified by the conditions shown in Table~\ref{tab:Intro:cut_condition}. 
\begin{table}[htbp]
\small{
\begin{center}
\begin{tabular}{|l|l|l|}  
\hline
                & Category   & Condition  (Backward $e^-$ emission case) \\
 \hline 
 \hline
  C0 &Prompt signal timing & $[1.5{\rm \mu s} <t_P< 5.5{\rm \mu s}]$\\
  \hline
   C1 &Prompt event topology &
               $[{\rm B0 \land B1 \land B2 \land \overline{F2}  }]$ \\
       &     &[Contained within one unit layer]\\
       &                 &[Single PANDA hit or two adjacent PANDA hits] \\
   \hline
   C2 &Consistency with MIP hits & [1MeV$< E_{\rm B1}, E_{\rm B2}<$5MeV],~~~[$0.5< (E_{\rm B1}/E_{\rm B2})<2$]\\
   \hline 
  C3 &$e^-$ visible energy &[15~MeV$<  E_P <$40~MeV]\\
  \hline
  \hline
    C4 & Delayed signal &[1.5~MeV$<  E_D <$ 9~MeV],~~~[$10{\rm \mu s} < t_D -t_P  <100{\rm \mu s}$]\\
  \hline
  \end{tabular}
\caption{\small{ $\nu_e$ event selection conditions. 
$t_P$ and $t_D$ are the time from the start of the first beam pulse for the prompt and delayed signals, respectively.  
$E_P$ and $E_D$ are sum of energies of all the scintillators contributing the signal in the prompt and delayed events, respectively.
 Event topology cuts C1 and C2 ensure  that minimum ionizing particle ($e^-$) is generated from the lead target.  
  Penetrating charged particles are removed by the anti-coincidence of F2.
For forward $e^-$ emission case, read the table by replacing Bn $\leftrightarrow$ Fn.
}}
\label{tab:Intro:cut_condition}
\end{center}
}
\end{table}
  C0 and C3 are the timing and energy window for prompt signal.

\begin{figure}[htbp]
\centering
\includegraphics[width=100mm] {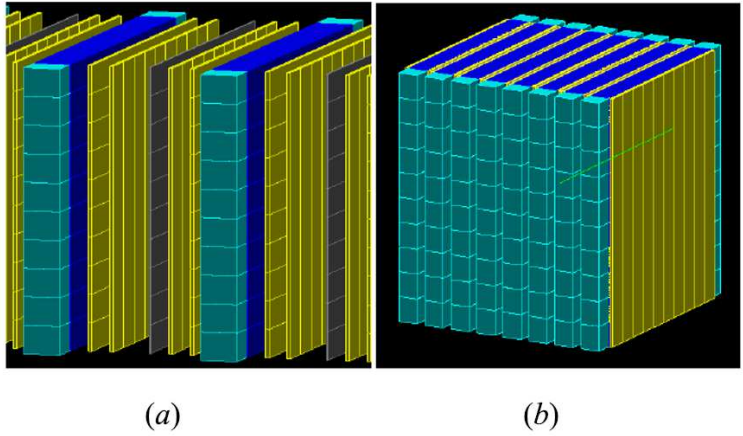}
\caption{\small{An example of the detector geometry (8$\times$10 case) formed in the Monte Carlo simulation. 
$(b)$  The overall view.
$(a)$  Take-apart view to see the detailed lead sandwich structure. 
Light blue cuboids show the PANDA modules. 
Yellow boards show the thin plastic scintillator.
Dark blue bands show the Gd sheets and black boards show lead boards.
}}
\label{fig:Intro:MC_geometry}
\end{figure}

\begin{figure}[htbp]
\centering
\includegraphics[width=140mm] {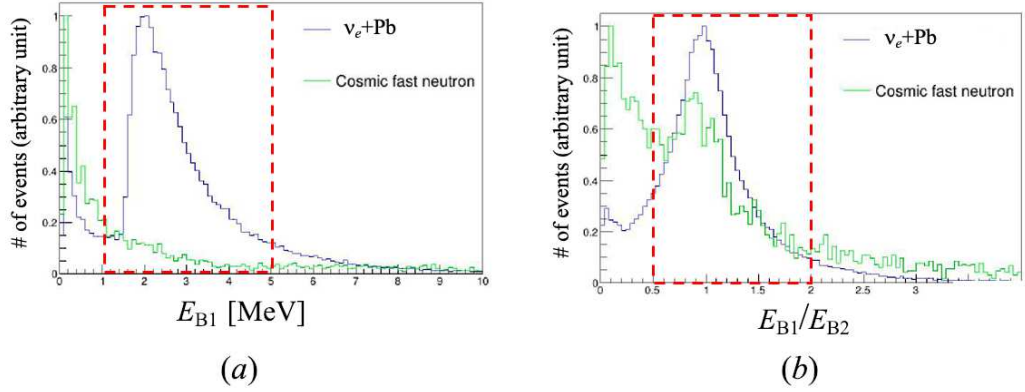}
\caption{\small{MC data. 
$(a)$ Energy deposit of scintillator strip B1; $E_{\rm B1}$. 
$(b)$ Ratio of B1 energy and B2 energy ($E_{\rm B1}/E_{\rm B2}$). 
The event selection regions are overlaid as red-dotted boxes.
}}
\label{fig:Intro:EB1_EB1EB2}
\end{figure}

We developed a Monte Carlo simulation of the DaRveX detector using GEANT4~\cite{GEANT4} and estimated event selection efficiencies and cosmic fast neutron rejection rate~\cite{Noguchi22}. 
Figs.~\ref{fig:Intro:MC_geometry} show  an examples of the detector geometry made in the simulation. 
We  also developed a simple $\nu_e$+Pb event generator.
The probability function of the electron energy is calculated by using the $\nu_e$ energy distribution; Eq.~(\ref{eq:App:hatGamma}), 
 $\nu_e+$Pb differential cross section; Fig.~\ref{fig:App:nu-Pb_Xsection} and Eq.~(\ref{eq:Ee=Enu-15MeV}). 
The emission angle distribution of the electron is assume to be backward peak (50~MeV line of Fig.~(\ref{fig:MainPart:e-_angle})) to check if directionality dependence can be measured. 
Figs.~\ref{fig:Intro:EB1_EB1EB2} show the simulated distribution of the cut parameters of C2. 
The cosmic-ray fast neutron events are also simulated by using existing simulation code~\cite{Hino21}.

\begin{figure}[htbp]
\centering
\includegraphics[width=70mm] {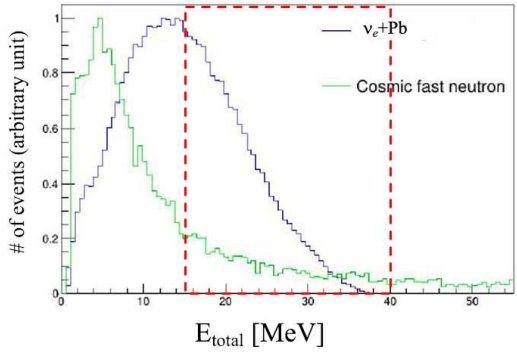}
\caption{\small{Distribution of the total energy deposited in the plastic scintillators.
The event selection window is overlaid as red-dot box.
The energy threshold (15~MeV) is to nearly maximize the signal to the cosmic-ray fast neutron ratio. 
}}
\label{fig:Intro:Etotal}
\end{figure}

Fig.~\ref{fig:Intro:Etotal} is the distribution of the sum of the energy deposited in scintillators (C3). 
If we compare the electron energy spectra of Fig.~\ref{fig:Intro:Etotal}  and yellow line of Fig.~\ref{fig:Intro:e-_energy_spectra}, we notice that electron loses $\sim 5$~MeV energy in the dead region of the detector such as lead board. 
This can be understood because if the electron is generated at the center of the lead board, it has to penetrate at least 2~mm of lead before it comes out and loses at least 4.5~MeV of energy within the lead board. 
The threshold energy, 15~MeV is to nearly maximize the S/N ratio in Fig.~\ref{fig:Intro:Etotal}.  
 
\begin{figure}[htbp]
\centering
\includegraphics[width=140mm] {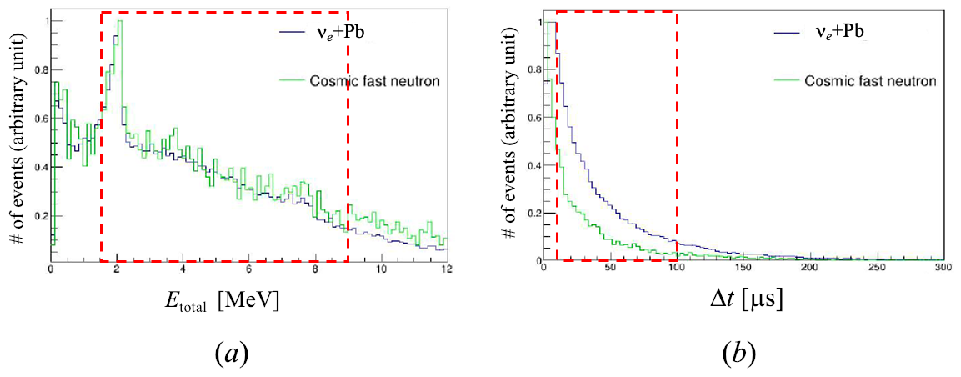}
\caption{\small{ Energy and timing distributions of the delayed neutron signal.  
$(a)$  Energy distribution. 
The peak at $E\sim$2~MeV is the $\gamma$-ray from neutron absorption to proton ($n+p \to d+ \gamma (2.2{\rm MeV})$). 
The small 8~MeV bump is the signal of neutron absorption to Gd ($n+{\rm Gd}\to {\rm Gd'} + \gamma$-rays ($\Sigma E_\gamma =8{\rm MeV})$). 
Gd emits several $\gamma$-rays and some of them  escape from the detector or lose energy in the lead board so that the visible energy of Gd signal becomes less than 8~MeV.
 $(b)$ Timing distribution of the delayed signal. 
 The MC simulation shows that the mean neutron absorption time is $\tau \sim 30\mu{\rm s}$. 
 The reference event selection regions are overlaid as red-dotted line boxes.
}}
\label{fig:Intro:delayed}
\end{figure}

 C4 is for delayed coincidence which removes single hit backgrounds.
Figs.~\ref{fig:Intro:delayed}$(a)$  show the visible energy and $(b)$ shows the timing distribution of the delayed signal. 
 From the energy distribution, the energy window for the delayed signal is set to be $1.5{\rm MeV} < E_D < 9{\rm MeV}$ to include neutron absorption by proton events. 
 From the timing distribution, the mean neutron absorption time is $\tau \sim 30{\rm \mu s}$ and the upper limit of the timing is chosen as $\Delta t < 3 \tau \sim 100\mu$s. 
 The shorter limit of the time difference cut of C4 (10~$\mu$s) is to remove the Michel electron, 
 \begin{equation}
  \mu^+ ({\rm stop~in~detector}) \to e^+ + \nu_e + \overline{\nu}_\mu~.
  \label{eq:Michel}
 \end{equation}
 The stopping $\mu^+$ may mimic the prompt signal and the produced $e^+$ in the decay of  $\mu^+$ may mimic the delayed neutron signal. 
By opening the delayed timing window $10~\mu$s after the prompt signal, 99\% of this correlated background can be removed ($e^{-(10\mu s/\tau_\mu )}=0.01$). 

The detection efficiencies are estimated by using the MC simulation and 
 Table~\ref{tab:Intro:cut_efficiency} summarizes the results.
\begin{table}[htbp]
\bigskip
\begin{center}
\begin{tabular}{|l||c|c|}  
\hline
           &   \multicolumn{2}{c|}{Efficiency}\\

\hline                       
          Cut          &   $\nu_e$  & Cosmic  fast neutron\\
 \hline 
 \hline
  \hline
 C0: Prompt timing ($\epsilon_{bt}$)     & 0.606   & $1\times10^{-4}$\\
 \hline
 \hline 
 C1$\times$C2: Event Topology ($\epsilon_{\rm ET}$)    &  0.36 & $9\times 10^{-3}$ \\
 \hline
 C3: $e^-$ visible Energy  ($\epsilon_{E} $) &   0.30  & 0.073 \\ 
 \hline  
  C4: Delayed Coincidence ($\epsilon_{\rm DC} $)  & 0.32 & 0.21\\ 
 \hline
  \hline
 C1$\sim$C4  & 0.062 &$3.2\times 10^{-4}$\\
\hline
\hline
 \hline
 Total  ($\epsilon_{\nu_e}$) & 0.037  & $3.2\times 10^{-8}$\\
   \hline
  \end{tabular}
\caption{\small{
$\nu_e$ selection efficiencies.  
C0 is from numerical calculation (Eq.~(\ref{eq:Intro:et})). 
C1$\sim$C4 are calculated by the MC simulation.
C0 and C1$\sim$C4 are independent and  $\epsilon_{\nu_e}$ is the product of them. 
}}
\label{tab:Intro:cut_efficiency}
\end{center}
\end{table}
The event selection efficiency of C1$\sim$C4 for $\nu_e$ is 6.2\%.
Together with the timing cut C0,  the total $\nu_e$ efficiency is 3.7\%. 
Assuming the beam power; $P_{\rm B}$=0.8~MW, baseline $L_B$=10~m and the target lead mass 
$M_{\rm Pb}$=250~kg, from Eq.~(\ref{eq:n_nuPb=}), the expected $\nu_e$ event rate is, 
\begin{equation}
  N_{\nu_e} = 9.6/{\rm day} \times 0.037= {\rm 0.36~event/day}~.
  \label{eq:intro:N_nue=}
\end{equation}

Assuming 200 days/year of beam time and 70\% of beam and data taking efficiency, we expect to measure 100 $\nu_e$+Pb events in 2 years of data taking.

 \section{Backgrounds}
 \label{sec:BKG}
 
 The main issue to realize the experiment is the background control.  
 Since the neutrino cross section is extremely small, it is necessary to reduce the background strongly by requiring various conditions for $\nu_e$ event selection. 
 In order to reduce background, the detector will be surrounded by cosmic-ray muon anti-counters. 
 Lead bricks and boronated rubber sheets will be placed around the detector to shield the $\gamma$-ray and thermal neutron backgrounds.

 \subsection{Origins of the Backgrounds}
  \label{sec:BKG_Origin}
%
There are two kinds of backgrounds for the $\nu_e+$Pb measurement.
One is the nature origin and the other is the beam origin. 
Table~\ref{tab:Intro:Natural_BKG} categorizes the possible backgrounds. 
\begin{table}[htbp]
\small{
\begin{center}
\begin{tabular}{|l|l||l|c|}  
\hline
           &            & Prompt Signal   & Delayed Signal       \\
 \cline{3-4}         
   Origin   &  Category      & ($t_P= 1.5 \sim  5.5\mu s$)   & ($t_D-t_P  = 10\sim 100\mu s$)    \\
        &         & ($E_P =15\sim  40$~MeV)      & ($E_D= 1.5 \sim  9$~MeV)    \\
 \cline{3-4}
    $p/\mu +N$         &         & $\epsilon_{bt} = 1\times 10^{-4}$    & $\epsilon_{bt} = 2.25\times 10^{-3}$   \\
 \hline
 \hline      
     &    [I] Thermal $n$     &      $n+A\to A^*\to \gamma$ &  
  $n+A\to A^*\to \textcolor{red}{\gamma}$ \\ 
 \cline{4-4}
      &  ($E_n \sim 25$~meV)             &                    &  
      $n+{\rm Gd} \to\textcolor{red}{ \gamma s} $  \\ 
               \cline{2-4}  
               &   [II]  Fast $n$   &  $n+p \to n+\textcolor{red}{p}$ 
  &  $n+{\rm Gd}\to \textcolor{red}{\gamma s}$ \\ 
  $  ~~~~\to n+X $ &  ($E_n > 20$~MeV) &  &  \\ 
\cline{2-4}
     &  [III] ($E_n>200$~MeV)   & &  \\ 
      &$n +A $  & $\mu^+ \to \textcolor{red}{e^+}$ & 
      $n+{\rm Gd} \to\textcolor{red}{ \gamma s} $   \\ 
      & ~~$\to (n, \pi^+) \to (n, \mu^+)$  &  &  \\ 
\hline
  $  ~~\to \pi^+ \to \mu^+ $ &[IV]  Michel electron 
  &   $\pi^+ \to \textcolor{red}{\mu^+}$ &  $\mu^+ \to \textcolor{red}{e^+}$ \\ 
  \hline
   $~~\to \pi^0 \to \gamma \gamma $ & [V] High energy $\gamma$ & 
    $\textcolor{red}{\gamma}$ &  $\textcolor{red}{\gamma}$ \\ 
  \hline
 ${\rm ^{238}U, ^{242}Th, ^{40}K}$ & [VI] $E_\gamma<$2.8~MeV & $\gamma$ &  $\textcolor{red}{\gamma}$ \\ 
  \hline
 \end{tabular}
\caption{\small{Background category.  
The particles written with red color generate the background signal.
 Cosmic-ray origin backgrounds are [I]$\sim$[V]. 
 Beam origin backgrounds are [I] $\sim$[III].
[II]$\sim$[IV] are correlated backgrounds. 
$\epsilon_{bt}$ is reduction factor due to the timing cut for the nature-origin backgrounds. 
}}
\label{tab:Intro:Natural_BKG}
\end{center}
}
\end{table}
 Both the cosmic-ray and the beam  can produce neutrons and $\gamma$s with various energies.
 
 [I] Neutrons produced by proton beam or cosmic-rays can thermalize colliding with nuclei.  
 The thermal neutron can be absorbed by nucleus around the detector and the excited nucleus can emit $\gamma$-rays whose typical energy is 10~MeV or less. 
 For the neutron produced by the proton beam,  the kinetic energies of neutron which can enter the time window of the prompt ($t_P>1.5\mu$s) and delayed ($t_D > 11.5\mu$s) timing window at 10~m distance, are $E_n < 0.23$MeV and $E_n < 3.9$keV, respectively. 
 Therefore, the neutrons before thermalization can not satisfy the energy thresholds for the $\nu_e$ selection.
 On the other hand, once a neutron becomes thermalized it can survive several hundred micro seconds and if it is absorbed by a nucleus, the nucleus can emit $\gamma$-rays whose energy is as much as 10~MeV and it can easily satisfy the energy threshold for the delayed signal. 
This is expected to be the most sever backgrounds for the delayed signal. 

 [II] Neutrons with energy 20~MeV or more can produce prompt like signal by hitting proton in the detector.  
The neutron loses energy and thermalizes in the detector and be captured by Gd, producing delayed signal. 
Since one neutron produces both prompt and delayed signals, it is called correlated backgrounds. 

[III] If the energy of the neutron is 200~MeV or higher, it can produce $\pi^+$, colliding with the target material and $\pi^+$ decays and produces $\mu^+$. 
$\mu^+$ can stop in the detector and decays and produces $e^+$ a few $\mu$s after. 
Therefore, if the original neutron is produced by the beam, the $e^+$ signal comes within the prompt timing window. 
On the other hand, if the original neutron thermalizes in the detector and is absorbed by Gd, it produces correlated delayed signal. 
This kind of backgrounds can be removed if the $\pi^+$ production or decay signals are identified within the on bunch timing.
Therefore, it is important to measure the events not only within the DAR $\nu_e$ event window but also on bunch timing. 

[IV] If cosmic $\mu^+$ stops in the detector and decays there,  the $\mu^+$ may mimic the prompt signal and  $e^+$ produced in the decay may mimic the delayed signal. 
Therefore, the Michel decay event is categorized as correlated background. 
The starting time of the delayed signal (10$\mu$s) is set to remove this kind of backgrounds.
If necessary the starting time can be delayed further without affecting much to the $\nu_e$ efficiency.

 [V]  High energy $\gamma$ rays are coming from decay of $\pi^0$  produced by the cosmic rays. 
  The probability for the $\gamma$-ray to satisfy the prompt event topology is small and this kind of $\gamma$-ray is expected to contribute as a delayed signal coinciding with other prompt-mimicking background signal. 
  
 [VI] The energies of $\gamma$-rays  in the decay chains of natural backgrounds, $^{238}$U, $^{232}$Th, $^{40}$K are low enough ($<$2.8~MeV) to eliminate by the total energy cut.
This kind of background can be reduced by lead shield. 

 \subsection{Cosmic-ray fast neutron}
  \label{sec:Cosmic-ray_fast_neutron}
%
The correlated background caused by the cosmic-ray fast neutron (CFN) is assumed to be one of the most sever backgrounds. 
The efficiencies of the CFN for the $\nu_e$ selection are calculated by the MC simulation and 
its values are shown in Table~\ref{tab:Intro:cut_efficiency}.
The cosmic fast neutron flux with energy $>1$~MeV in a building at Tokyo is roughly $\sim 10[{\rm /m^2 s}] $\cite{Takayasu09}.
Since effective cross section of the DaRveX detector is $\sim1~[{\rm m}^2]$, the expected CFN rate is,  
\begin{equation}
 n_{\rm CFN} \sim 10[{\rm /m^2 s}] \times 1 [{\rm m^2}] \times (3.2 \times 10^{-8})  
\times (8.64\times 10^4 [{\rm s/day}] ) \sim 0.028[{\rm /day}]
\end{equation}
This is less than 1/10 of  the expected $\nu_e$ rate: Eq.~(\ref{eq:intro:N_nue=}). 
However, this is  a very rough estimation and we are planning to measure the CFN background rate directly with a prototype detector. 

 \subsection{Beam origin background}
  \label{sec:Beam_origin_bckground}
 The rate of the beam origin backgrounds strongly depend on the experimental site conditions. 
 In order to measure the properties of such kind of back grounds, we performed on-site background measurement that is explained in Appendix-3. 

\section{Cross section measurement}
The  $\nu_e + {\rm Pb}$ cross section is calculated as 
\begin{equation}
 \sigma_{\nu_e {\rm Pb}}= \frac{n_{\nu_e}}{A_{\rm Pb} f_{\nu_e} \epsilon_{\nu_e}}~.
\end{equation}
The $\nu_e$ flux will be measured by JSNS$^2$ experiment using $\nu_e + {\rm ^{12}C} \to e^- + {\rm ^{12}N_{gs}}$ events with precision $\delta  f_{\nu_e} / f_{\nu_e}  \sim 10\%$~\cite{JSNS2-Proposal13}.
As estimated in Section~\ref{sec:main:event_selection}, the number of events to be obtained 
with  0.8~MW of beam power and 2 years of data taking, we expect to obtain $N_{\nu_e}\sim$ 100 events.
We aim to achieve the signal to noise ratio; S/N=1, that is the same amount of backgrounds ($N_{\rm BG} \sim 100$) are included in the event sample 
($n_{\rm sample}\sim 200$). 
Therefore, the statistical error is $\delta N_{\nu_e} /  N_{\nu_e} \sim 14\%$. 
The background contamination fraction can be estimated statistically by the  information of the time distribution and the event topology. 
Assuming the systematic error of the event selection efficiency and background fraction estimation is 
$\sigma_{\rm sys} \sim 10\%$, 
the error of the cross section measurement is expected to be 
$\delta \sigma_{\rm \nu_e Pb} /  \sigma_{\rm \nu_e Pb} \sim 20\%$. 
If the actual  $\nu_e$+Pb cross section is unexpectedly small and the observed events are consistent with the background-only hypothesis, 
the statistic error is 10\%. 
Together with the $\nu_e$ flux and the systematic uncertainties, 
the corresponding  2$\sigma$ upper limit of the $\nu_e^\mu$-spectrum weighted total cross section (Eq.~(\ref{eq:App:<sigma>})) is, 
\begin{equation}
 \braket{\sigma_{\nu_e^\mu {\rm Pb}}}< 2 \sqrt{0.03}\times 3.75\times 10^{-39}[{\rm cm^2}] = 1.3\times 10^{-39}[{\rm cm^2}]~. 
\end{equation}

The error can be reduced by extending the run time and understanding the systematic uncertainty with real data.

\section{Summary}
A conceptual design of experiment for decay at rest $\nu_e +{\rm Pb} \to  e^- + xn + {\rm Bi} $  cross section measurement at J-PARC MLF is presented. 
MLF provides a powerful $\nu_e$ beam with very narrow pulses and it is world's most suitable accelerator for this kind of measurement. 

The neutrino detector is designed by remodeling existing PANDA reactor neutrino detector. 
A 4~mm thick lead board and four 1~cm thick plastic scintillators are sandwiched between PANDA 
 plastic scintillator blocks. 
Not only the absolute cross section, but also energy and direction of the final state $e^-$ can be measured in this experiment.
By making use of the beam timing,  topological information of the events and delayed coincidence, and strong neutron  and 
$\gamma$-ray shields, backgrounds are expected to be strongly suppressed. 
The expected number of events is 100 with 2~years of running at 10~m from the Hg target. 
The expected error of the cross section measurement is $\sim$20~\%.
This differential cross section provides useful information to understand the  neutrino-nucleus interactions.
It can also be used in low energy $\nu_e$ related experiments such as low energy $\nu_e$ oscillation measurements and  flavor specific detection of supernova explosion $\nu_e$. 
If anisotropy of $e^-$ emission angle distribution is observed in this experiment, measurement of the supernova direction will become possible from the measurement of $\nu_e$.

\section{Appendix-1: Signal property}
\label{sec:App1}
 In the following subsections, the $\nu_e$+Pb reaction rate is estimated using the reference condition with the parameters shown in Table-\ref{tab:Signal:Reference_Value}.
\begin{table}[htbp]
\begin{center}
\begin{tabular}{|l||c|c|c|c|}  
\hline
 Parameter    & Beam  & Proton  & Baseline & Lead \\
            & power & energy &     & weight  \\
 \hline
  Symbol   &  $P_{\rm B}^*$ &  $E_{\rm p}^*$ & $L_{\rm B}^*$ & $M_{\rm Pb}^*$ \\ 
\hline \hline
  Reference value   & 1[MW]  & 3[GeV]     & 10[m] & 1[ton] \\
\hline
 \end{tabular}
\caption{\small{
Parameter values of the reference condition used to estimate the $\nu_e$+Pb reaction rate.
}}
\label{tab:Signal:Reference_Value}
\end{center}
\end{table}
%

 \subsection{$\nu_e$ flux }
 
\begin{figure}[htbp]
\centering
\includegraphics[width=150mm]
{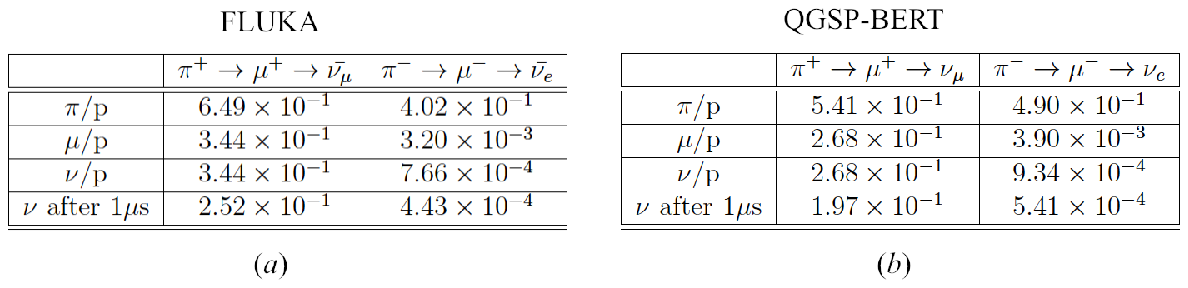}
\caption{\small{
Estimations of $\mu$ DAR neutrino production by 3~GeV protons using $(a)$ FLUKA and $(b)$ QGSP-BERT hadron simulation packages~\cite{JSNS2-TDR17}.  
There is 25\% discrepancy for $\nu/p$ value. 
The average value ($\nu /{\rm p}=0.306$) is used in this study.  
}}
\label{fig:nue_per_p}
\end{figure}

The number of $\overline{\nu}_\mu$ per beam proton at J-PARC MLF was estimated in the JSNS$^2$ proposal
\cite{JSNS2-Proposal13} by using 
 FLUKA\cite{FLUKA} and QGSP-BERT simulations. 
 Since our $\nu_e$ is generated together with $\overline{\nu}_\mu$ in $\mu^+$ decay at rest (DAR), we can regard the number of $\nu_e$ produced is the same as the number of $\overline{\nu}_\mu$.  
Fig.~\ref{fig:nue_per_p} shows the simulated number of $\overline{\nu}_\mu$ per proton hit, copied from \cite{JSNS2-Proposal13}. 
There is 25\% discrepancy between the two simulations. 
By taking the average of the two results, we use  $\rho_\nu=\nu/p = 0.306$.
The neutrino generation rate for the reference operation condition is,
\begin{equation}
 G_\nu^* =  \rho_\nu\frac{P_{\rm B}^*}{E_{\rm p}^*} 
 = 0.306[\nu /p]
 \frac{10^6[{\rm J/s}]}{(3\times 10^9[{\rm eV}/p]) \times (1.6 \times 10^{-19}[{\rm J/eV}])}
 =6.38\times 10^{14} [{\rm \nu / s}]~.
\end{equation}
When $\pi^+$ stops in the target, the directional information is completely lost and the neutrino is emitted isotropically. 
The reference $\nu_e$ flux ($f_\nu^*$)  is then, 
\begin{equation}
 f_\nu^*= \frac{G_\nu^*}{4\pi L_{\rm B}^{*2}}
 = \frac{6.38\times 10^{14} [{\rm \nu / s}]}{4\pi(10[{\rm m}])^2}
 =5.08\times 10^7 \left[\frac{\nu}{\rm s\cdot cm^2 } \right]~.
 \label{eq:reffnu}
\end{equation}

\subsection{$\nu_e^\mu$ energy spectrum}
We will consider possible neutrino oscillations of the neutrinos produced in $\pi^+$ and $\mu^+$ DAR as shown in 
Fig.~\ref{fig:intro:DaR_nu_Oscillation}. 
In such case there are two kinds of $\nu_e$, one is produced in the decay of $\mu^+$ and another one is produced in the oscillation of $\nu_\mu$ which is produced in $\pi^+$ decay. 
We call the former as $\nu_e^\mu$ and the latter as $\nu_e^\pi$. 
 
The $\nu_e^\mu$ energy spectrum can be calculated precisely by using the electroweak theory. 
The normalized energy spectrum of $\nu_e$ in $\mu^+ \to e^+ + \nu_e + \overline{\nu}_\mu$ DAR is~\cite{Suekane1500},
\begin{equation}
 \hat{\Gamma}_{\nu_e^\mu } (E_\nu )
 = \frac{96}{m_\mu} \left(\frac{E_\nu}{m_\mu}\right)^2 \left( 1-2\frac{E_\nu}{m_\mu} \right)
 =0.908 \left(\frac{E_\nu}{m_\mu}\right)^2 \left( 1-2\frac{E_\nu}{m_\mu} \right) [{\rm /MeV}]~,
 \label{eq:App:hatGamma}
\end{equation}
where $0<E_\nu< m_\mu/2 = 53$~MeV.
The shape of the energy spectrum is shown in Fig.~\ref{fig:intro:DaR_nu_spectra} as the green line or in  Fig.~\ref{fig:SN_neutrino} as the red dotted line. 

The average $\nu_e^\mu$ energy  is
\begin{equation}
 \braket{E_{\nu_e^\mu}} =\int_0^{m_\mu/2}E_\nu \hat{\Gamma}_{\nu_e^\mu } (E_\nu )dE_\nu
 =\frac{3}{10}m_\mu = 31.8~{\rm MeV}~, 
\end{equation}
which is close to the monochromatic energy of $\nu_e^\pi$ (30~MeV). 
 Together with the reference neutrino flux (\ref{eq:reffnu}), the  DAR reference $\nu_e^\mu$ differential flux is
 \begin{equation}
  F_{\nu_e^\mu}^* (E_\nu)= f_\nu^*  \hat{\Gamma}_{\nu_e^\mu}(E_\nu)
  =4.60\times 10^7   \left(\frac{E_\nu}{m_\mu}\right)^2 \left( 1-2\frac{E_\nu}{m_\mu} \right)\left[\frac{\nu_e}{\rm s\cdot cm^2 \cdot MeV} \right]
 \end{equation}
 
\subsection{$\nu_e+{\rm Pb} \to e^- + xn +{\rm Bi}$ cross sections}
\begin{figure}[htbp]
\centering
\includegraphics[width=12cm]{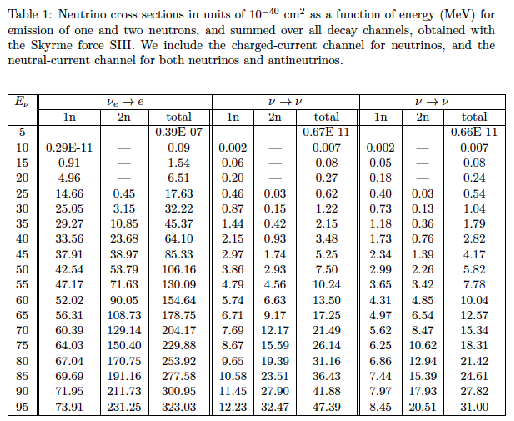}
\caption{\small{
$\nu$-Pb cross sections and caption taken from~\cite{Engel18}.
The neutral current cross sections are much smaller ($\sim 5\%$@40MeV) than the charged current cross section. 
}}
\label{fig:App:nu-Pb_Xsection}
\end{figure}

We will detect the electron and neutron(s) produced in the following reaction and measure the differential cross section with respect to the electron  energy and forward-backward asymmetry of the electron. 
\begin{equation}
 \nu_e+{\rm Pb} \to e^- + xn +{\rm Bi};~~~~~~(x=1{\rm ~or~}2)
 \label{eq:nue+PB->xn+Bi}
\end{equation}
The $\nu_e$ energy dependent cross sections ($\sigma_{\nu_e {\rm Pb}}$)  of the reaction (\ref{eq:nue+PB->xn+Bi}) are calculated in \cite{Engel18} 
which are shown in the table of Fig.~\ref{fig:App:nu-Pb_Xsection} and we use the results of the calculation as a reference for our estimations.
\begin{figure}[htbp]
\centering
\includegraphics[width=100mm]{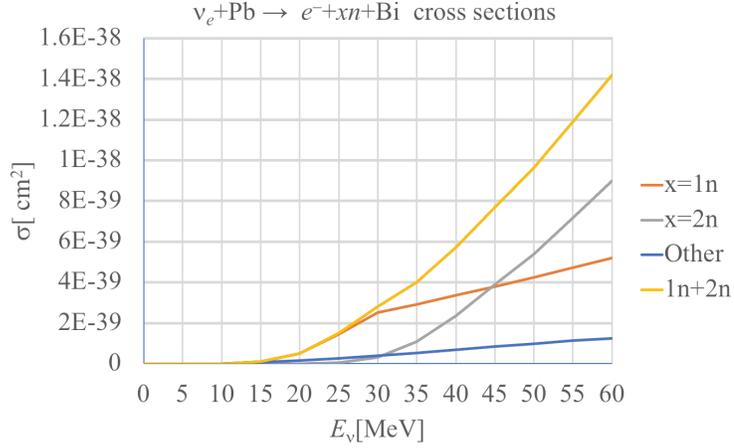}
\caption{\small{
The cross section of $\nu_e + {\rm Pb} \to e^- + xn + {\rm Bi}$ reactions from the table shown in Fig.~\ref{fig:App:nu-Pb_Xsection}. 
"Other" means $\sigma_{\rm Other}=\sigma_{\rm total} - \sigma_{2n} - \sigma_{1n}$.
}}
\label{fig:G*sigma}
\end{figure}
They calculated the cross section for the final states $1n+{\rm Bi}$ and $2n+ {\rm Bi}$, separately. 
Fig.~\ref{fig:G*sigma} shows the neutrino energy dependence of the $\sigma_{\nu_e {\rm Pb}}$ for $x$=1 or 2 and the sum of them. 
The "Other" means $\sigma_{\rm Other}=\sigma_{\rm total} -\sigma_{1n}- \sigma_{2n}$. 
$\sigma_{1n}$ and $\sigma_{2n}$ start to rise at around 15~MeV and 25~MeV, respectively because in order to emit a neutron, $\sim$10~MeV of energy is necessary to overcome the nuclear binding energy. 
Although energy dependence of the $\sigma_{1n}$ has a kink at $E_\nu$=30~MeV, the total neutron emission cross section, 
$\sigma_{1n} + \sigma_{2n}$ shows a monotonous rise. 

\begin{figure}[htbp]
\centering
\includegraphics[width=100mm]{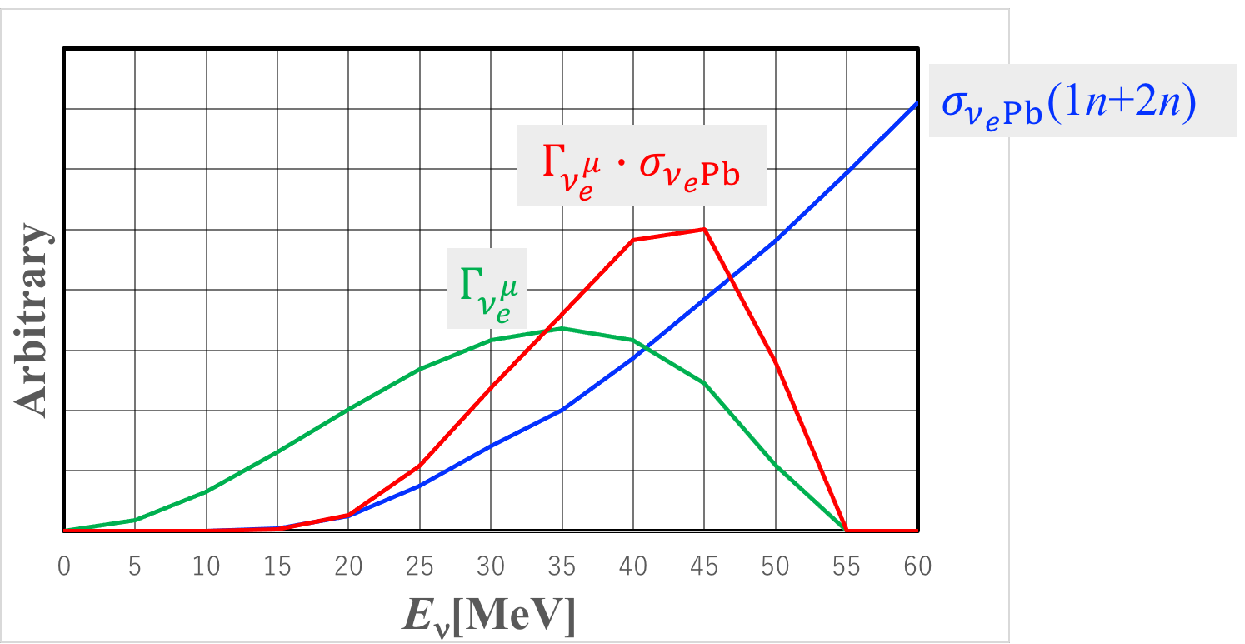}
\caption{\small{
Green: The energy spectrum of  $\nu_e^\mu$; $\Gamma_{\nu_e^\mu}$. 
Blue: $\nu_e + {\rm Pb} \to e^- + xn + {\rm Bi}$ cross section; $\sigma_{\nu_e {\rm Pb}}$ for $x$=1 or 2. 
Red: The energy spectrum of the neutrino which performs the $\nu_e + {\rm Pb}$ reaction; 
$\sigma_{\nu_e {\rm Pb}}\Gamma_{\nu_e^\mu}$.
The scale of the vertical axis is arbitrary. 
}}
\label{fig:Flux_Xsection_FX}
\end{figure}

Fig.~\ref{fig:Flux_Xsection_FX} shows the $\nu_e^\mu$ energy spectrum which performs the reaction~(\ref{eq:nue+PB->xn+Bi}) calculated by the products of the $\nu_e^\mu$ energy spectrum and neutron emission cross sections, $\sigma_{1n}+\sigma_{2n}$.
Since the cross section increases according to the energy, the mean energy of the neutrinos which participate in the reaction is larger than that of the original neutrino spectrum. 
The mean energy of the original neutrino spectrum is  32~MeV, while the mean energy of the neutrinos which performed the reaction is 39~MeV. 

The $\nu_e^\mu$-spectrum weighted total ($1n$+$2n$) cross section is
\begin{equation}
 \braket{\sigma_{\nu_e^\mu {\rm Pb}}} = \int \hat{\Gamma}_{\nu_e^\mu}(E_\nu)\times \sigma_{\nu_e {\rm Pb}}(E_\nu) dE_\nu 
 = 3.75\times 10^{-39}[{\rm cm^2}]
 \label{eq:App:<sigma>}
\end{equation}
 The event rate with the reference parameters can be calculated by multiplying the reference neutrino flux 
 $f_\nu^* [{\rm /cm^2 /s}]$, the number of target nuclei $A_{\rm Pb}^*$ and $\braket{\sigma_{\nu_e^\mu {\rm Pb}}}$. 
 
The number of Pb nuclei in the reference mass $M_{\rm Pb}^* =1$~ton is,
\begin{equation}
  A_{\rm Pb}^*= 10^6[{\rm g}]\times (6.02\times 10^{23}/207) [\rm Pb /g] = 2.91\times 10^{27}[{\rm Pb}]
\end{equation}
Therefore, the reaction rate for the reference parameters (Table-\ref{tab:Signal:Reference_Value}) is
\begin{equation}
 \begin{split}
 n_{\rm \nu_e^\mu Pb}^*&\sim A_{\rm Pb}^* f_\nu^* \braket{\sigma_{\nu_e^\mu {\rm Pb}}} \\
& =(2.91\times 10^{27}[{\rm Pb}]) \times \left(5.08\times 10^7 \left[\frac{\nu_e}{\rm s\cdot cm^2} \right]\right) \times (3.75\times 10^{-39}[{\rm cm^2 / Pb}]) \\
 &= 5.54\times 10^{-4}[\nu_e/{\rm s}] = \underline{47.9[\nu_e/{\rm day}] }
 \end{split}
 \label{eq:Intro:hat(nnue)}
\end{equation}

The isotope dependence of the cross sections  is expected to be small \cite{Engel18}  and we ignore the difference between the lead isotopes in this report. 
 
 \subsubsection{$e^-$ energy spectrum}
 In order to evaluate the detection efficiency of the $\nu_e + {\rm Pb}$ events, it is necessary to know the energy spectrum of the emitted electron. 
 However,  it is not written in reference~\cite{Engel18}.
  Therefore, we approximated the $e^-$ energy spectrum as follows.

 The kinetic energy of $e^-$ in the reactions, 
 \begin{equation}
   \begin{cases}
     \nu_e +{\rm ^{208}Pb} \to e^- + ~n + {\rm ^{207}Bi^+} \\
     \nu_e +{\rm  ^{208}Pb} \to e^- + 2n + {\rm ^{206}Bi^+}
   \end{cases}
   \label{eq:Signal:nue+208Pb->e-+n+207Bi}
 \end{equation}
 is the $\nu_e$ energy minus the energy transferred to nucleus, minus the electron mass $m_e$. 
 The energy transferred to nucleus is the mass difference between the initial state nucleus and final state nucleus + mass and kinetic energy of the emitted neutron(s).
  In order to calculate the mass difference between initial state and final state nucleus, their mass excesses are taken from the "Table of Isotopes" as in Table-\ref{tab:Mass_excess}. 
 \begin{table}[htbp]
\begin{center}
\begin{tabular}{|l||c|c|c|c|c|}  
\hline
 & $n$ & ${\rm ^{208}Pb }$ & ${\rm ^{208}Bi }$ & ${\rm ^{207}Bi}$ & ${\rm ^{206}Bi}$ \\
 \hline
  Mass excess $\Delta$ [MeV]   & $8.071$  & $-21.764$ & $-18.884$  & $-20.068$ & $-20.043$ \\ 
\hline 
 \end{tabular}
\caption{\small{
The mass excess of the atom related to the reaction (\ref{eq:nue+PB->xn+Bi}).
From "Table of Isotopes" \cite{TOI96}. 
}}
\label{tab:Mass_excess}
\end{center}
\end{table}
The mass excess of the atom with atomic number $Z$ and mass number $A$ is defined as 
\begin{equation}
 \Delta[Z,A] \equiv M[Z,A] -A u
\end{equation}
where $M$ is the mass of the atom and $u$ is the atomic mass unit defined by 
\begin{equation}
 u \equiv \frac{M[{\rm ^{12}C}]}{12} = 931.494~102~{\rm MeV}~. 
\end{equation}
The difference of sum of the mass between the initial and final states of the reactions (\ref{eq:Signal:nue+208Pb->e-+n+207Bi}) can be calculated as follows. 
\begin{equation}
 \begin{cases}
 \Delta E_{0n} = \Delta [^{208}{\rm Pb}] - \Delta[{^{208}{\rm Bi}}] = -2.9~{\rm MeV} \\
 \Delta E_{1n} = \Delta [^{208}{\rm Pb}] - \Delta[{^{207}{\rm Bi}}] -~\Delta[n]= -9.8~{\rm MeV} \\
 \Delta E_{2n} = \Delta [^{208}{\rm Pb}] - \Delta[{^{206}{\rm Bi}}] -2\Delta[n] = -17.9~{\rm MeV}
 \end{cases}
 \label{eq:Signal:DM=D208-D207-Dn}
\end{equation}
They mean that in order to emit one (two) neutron(s), 9.8(17.9)~MeV is necessary to overcome the potential energy of the nucleus.

\begin{figure}[htbp]
\centering
\includegraphics[width=80mm]{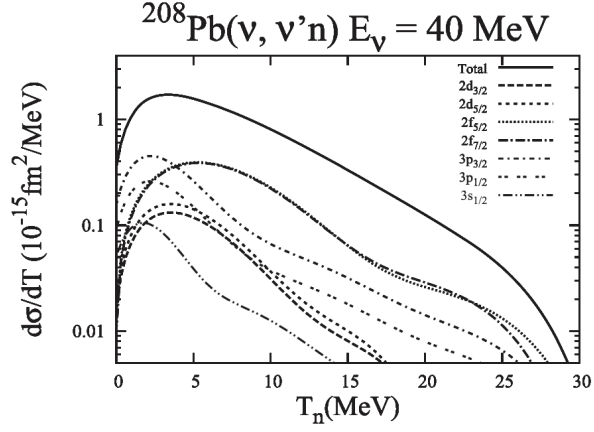}
\caption{\small{Neutron kinetic energy emitted by the neutral current interaction with 40~MeV incident neutrino
\cite{Hedayatipoor18}. 
}}
\label{fig:Signal:Tn}
\end{figure}

\begin{figure}[htbp]
\centering
\includegraphics[width=140mm]{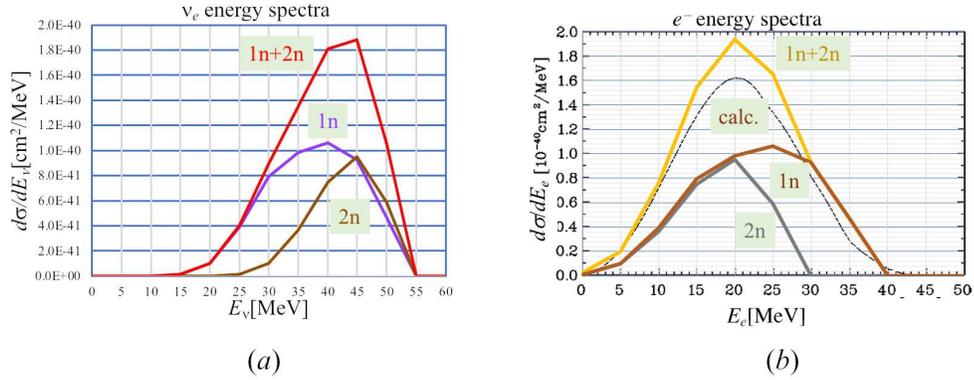}
\caption{\small{ $(a)$ Energy distribution of $\nu_e$ which emits one or two neutrons in the reaction $\nu_e + {\rm Pb} \to e^- + xn +{\rm Bi}$. 
The average 1$n$-$\nu_e$ energy is  37~MeV and that of 2$n$-$\nu_e$ is 43~MeV and that of total is 39~MeV. 
$(b)$ Solid lines are expected energy distributions of $e^-$ emitted in the reaction associated with 1$n$ or 2$n$ emissions. 
These spectra are obtained by subtracting 15~MeV from 1$n$-$\nu_e$ spectrum and 25~MeV from 2$n$-$\nu_e$ spectrum. 
The dashed line shows a calculated $e^-$ energy spectrum taken from \cite{Kolbe01}.
 The agreement between our $e^-$ energy spectrum for $1n+2n$ mode and the calculation is reasonably good.
}}
\label{fig:APP1:nue_e-_spectra}
\end{figure}

Fig.~\ref{fig:Signal:Tn} shows the kinetic energy of the neutron emitted by the neutral current interaction with 40~MeV incident neutrino \cite{Hedayatipoor18}. 
Roughly speaking, a neutron takes out $\sim$5~MeV as its kinetic energy.
Assuming it is not significantly different for NC and CC cases, we approximate the kinetic energy of the final state electron in the reaction~(\ref{eq:nue+PB->xn+Bi}) as, 
\begin{equation}
  E_e\sim 
  \begin{cases}
  E_{\nu_e}-15~{\rm MeV~~for}~1n~{\rm emission~events} \\
  E_{\nu_e}-25~{\rm MeV~~for}~2n~{\rm emission~events}
  \end{cases}
  \label{eq:Ee=Enu-15MeV}
\end{equation}
in order to use to estimate the detection efficiency, etc. 
The energy thresholds shown in the table of Fig.~\ref{fig:App:nu-Pb_Xsection} are consistent with this assumption. 
Fig.~\ref{fig:APP1:nue_e-_spectra}$(a)$ shows the energy distribution of $\nu_e$ for 1$n$ and 2$n$ emission cases.  
The average $\nu_e$ energy which contributes to $1n$ emission is  37~MeV and that of 2$n$ emission is 43~MeV. 
The difference between the average $\nu_e$ energies contributing to $1n$ and $2n$ emissions can be used to measure distortion of $\nu_e$ energy spectrum due to neutrino oscillation in future experiments. 
The solid lines in Fig.~\ref{fig:APP1:nue_e-_spectra}$(b)$ show the expected energy distribution of $e^-$ associated with the $1n$ and $2n$ emission reactions obtained by using Eq.~(\ref{eq:Ee=Enu-15MeV}).  
The electron energy peaks at around $E_e\sim 20$~MeV and the spectrum has symmetric feet. 
The dashed line in Fig.~\ref{fig:APP1:nue_e-_spectra}$(b)$ is a calculated $e^-$ energy spectrum taken from \cite{Kolbe01}.
The agreement is good despite the roughness of our approximations.
Especially the agreement of the spectrum shape, which is important to estimate the event selection efficiency, is quite good. 
%

\section{Appendix-2: Future scientific potentials}
\label{sec:App2}
\subsection{Neutrino oscillation measurements with decay at rest $\nu_e$}
\label{sec:App2:DAR}

Fig.~\ref{fig:intro:DaR_nu_Oscillation} shows possible neutrino oscillation measurements with DAR neutrinos, enabled by detecting low energy $\nu_e$. 
In this subsection, statistical sensitivities for DAR $\nu_e$ appearance and disappearance oscillation measurements are calculated based on simple assumptions. 

The disappearance $\nu_e \to \nu_e$ oscillation  is CPT inverted process of $\overline{\nu}_e \to \overline{\nu}_e$ oscillation and both  probabilities are exactly the same due to the CPT symmetry.
From measurement of baseline dependence of the disappearance of $\nu_e^\mu \to \nu_e^\mu$, at baselines of a few tens of meters,  tests of the reactor neutrino anomaly become possible.  

On the other hand, the appearance of $\nu_\mu \to \nu_e$ oscillation is the CP inverted process of $\overline{\nu}_\mu \to \overline{\nu}_e$ oscillation.
There is a possibility that CP symmetry is violated and the oscillation probabilities of $\nu_\mu \to \nu_e$ and $\overline{\nu}_\mu \to \overline{\nu}_e$ are not necessarily the same and both measurements are important. 
 By observing $\nu_\mu \to \nu_e^{\pi}$ it is possible to perform new sterile neutrino search  and a clean CP violation measurement\footnote{Clean CPV measurement with $\Delta m_{13}^2$ oscillation using DAR neutrinos is discussed in \cite{Grassi18}.} comparing with the result of  $\overline{\nu}_\mu \to \overline{\nu}_e$ measurements by JSNS$^2$\cite{JSNS2-II_2012}. 

In this appendix, statistical sensitivities of the $\nu_e$ appearance and disappearance measurements are estimated assuming there is a  near detector at a baseline $L_N$=10~m with 11~ton lead target mass and 
a far detector at a baseline $L_F=$30~m with 100~ton lead target mass\footnote{
There exists 76~ton lead target supernova $\nu_e$ detector called HALO\cite{HALO15}. 
 Therefore, 100~ton lead target detector is not unrealistic scenario.  } .
 In such the Near/Far configuration, the neutrino flux times the number of target nuclei is identical for both the near and far detectors,
\begin{equation}
   \frac{G_\nu}{4\pi L_F^2} A_{\rm Pb }[L_F]=    \frac{G_\nu}{4\pi L_N^2} A_{\rm Pb }[L_N]\equiv \phi_{fA}
 \label{eq:App:NF=NN}
\end{equation}
and the expected number of events are the same for both near and far detectors, if there is no neutrino oscillation. 
 This condition makes the sensitivity analysis easier.
 If we assume to take data for 5~years (=1,000~data taking days) with 1~MW beam,  the expected number of $\nu_e^\mu$ reaction is, from Eq.~(\ref{eq:Intro:hat(nnue)}), 
 \begin{equation}
   N_{\nu_e^\mu}^0 = 47.9[{\rm \nu_e/day/ton}]\times 11[{\rm ton}]\times 1,000[{\rm day}] =530,000
   \label{eq:Nnuemu=}
 \end{equation}
 in each detector. 
 %
\subsubsection{Disappearance of $\nu_e^\mu$: direct test of reactor neutrino anomaly}

\begin{figure}[htbp]
\centering
 \includegraphics[width=12cm]{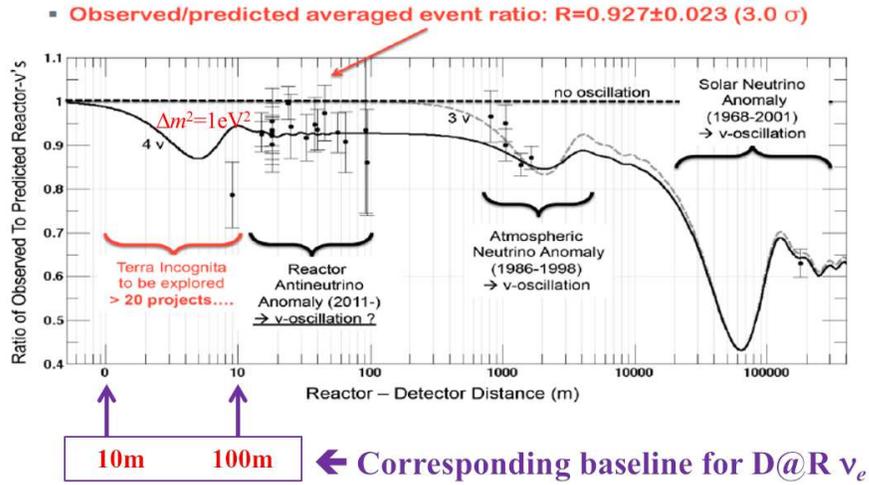}
\caption{\small{
Reactor neutrino anomaly \cite{Lasserre2012}. 
The observed number of $\overline{\nu}_e$ at baseline $L>10$~m is $\sim 7 \%$ smaller than expectation. 
This deficit of $\overline{\nu}_e$ can be explained if there is a sterile neutrino with mass $>$ 1~eV and mixing with active neutrino is $\sin^22\theta \sim 0.1$. 
Since energy of DAR $\nu_e^\mu$ is 10 times larger than that of reactor  $\overline{\nu}_e$, the corresponding oscillation length becomes 10 times longer and baseline dependence of the oscillation for $\Delta m^2 \sim 1 {\rm ~eV^2}$ can be measured.
}}
\label{fig:DaR_nu_Oscillation}
\end{figure}

Fig.~\ref{fig:DaR_nu_Oscillation} explains the reactor neutrino anomaly. 
The observed number of $\overline{\nu}_e$ is  $\sim 7 \%$ smaller than expectation at baseline $L>10$~m. 
This deficit of $\overline{\nu}_e$ can be explained if there is sterile neutrino with mass larger than 1~eV and mixing angle to the active neutrino is $\sin^22\theta \sim 0.1$. 
In such the case, the first oscillation maximum becomes $L < 5$~m for reactor neutrinos.  
However, it is difficult to measure the oscillation pattern at this baseline because the typical size of rector core is a few meter and the oscillation pattern is washed out because the baseline depends on where in the core the neutrino is generated. 
On the other hand, since the energy of DAR neutrino is 10 times larger than that of reactor neutrino, the first oscillation maximum becomes $\sim$50~m for neutrino with mass 1~eV, which is accessible if a few tens of tons of lead based detector is used. 

If neutrino oscillation exists,  the number of events to detect at far and near detectors is 
\begin{equation}
 n_{\nu_e^\mu}^D[L]= \epsilon_{\nu_e^\mu}^D \left(N_0 - \Lambda_{\rm O}[L] \sin^2 2\theta_D \right)
 \label{eq:App:N=e(N0-sinqL)}
\end{equation}
where, the sub- and super-script ``$D$,, represents the parameters for disappearance.  
$L=L_F$ is for the far detector and $L=L_N$ is for the near detector and $\epsilon_{\nu_e^\mu}^D$ is the detection efficiency. 
Because the detectors here are much larger than the DaRveX detector, we assume the $\nu_e$ detection efficiencies in table~\ref{tab:Intro:cut_efficiency} increases to 
\begin{equation}
   \epsilon_{\nu_e^\mu}^D \sim 10\%~. 
   \label{eq:App:epsilinnuemu}
 \end{equation}

$N_0$ is the total number of $\nu_e^\mu$+Pb reaction in case there is no oscillation, 
\begin{equation}
 N_0 =  T  \phi_{fA} \int \hat{\Gamma}_{\nu_e^\mu}[E_\nu] \sigma_{\rm \nu_e Pb}[E_\nu] dE_\nu 
 =T \phi_{fA} \braket{\sigma_{\rm \nu_e^\mu Pb}}~.
 \label{eq:App:N0=}
\end{equation}
where $T$ is the data taking period, 
$\hat{\Gamma}_{\nu_e^\mu}$ is the normalized $\nu_e^\mu$ energy spectrum shown in Eq.~(\ref{eq:App:hatGamma}), 
$\sigma_{\rm \nu_e Pb}$ is the $\nu_e$-Pb reaction cross section and
$\braket{\sigma_{\rm \nu_e^\mu Pb}}$ is $\nu_e^\mu$-spectrum weighted cross section defined by Eq.~(\ref{eq:App:<sigma>}).
$\Lambda_{\rm O}\sin^22\theta_D$ is the number of event disappeared due to the neutrino oscillation.
\begin{equation}
 \Lambda_{\rm O} [L] = T  \phi_{fA} \int \hat{\Gamma}_{\nu_e^\mu}[E_\nu] \sigma_{\rm \nu_e Pb}[E_\nu] \sin^2\Phi_D dE_\nu 
 \label{eq:App:Lambda=}
\end{equation}
$\Phi_D = \Delta m^2_D L/4E_\nu$ is the oscillation phase. 

By relating the number of events taken in the far and near detectors, the mixing angle can be obtained as follows.
\begin{equation}
 \sin^2 2\theta_D = 
  \frac{n_N-n_F}{n_N \hat{\Lambda}_{\rm O}[L_F] -n_F \hat{\Lambda}_{\rm O}[F_N]}
 \label{eq:Intro:sin22q=}
\end{equation}
where $n_F=n_{\nu_e^\mu}^D[L_F]$, $n_N=n_{\nu_e^\mu}^D[L_N]$ and 
\begin{equation}
 \hat{\Lambda}_{\rm O} [L]= \frac{\Lambda_{\rm O}[L]}{N_0} 
 = \int  \left( \frac{\sigma_{\rm \nu_e Pb}[E_\nu]}{\braket{\sigma_{\rm \nu_e^\mu Pb}}} \right) 
 \hat{\Gamma}_{\nu_e^\mu}[E_\nu] \sin^2\Phi_D dE_\nu .
 \label{eq:App:hatLambda=}
\end{equation}
Note that the parameters $\epsilon_{\nu_e^\mu}^D T \phi_{fA}$ and the absolute scale of $ \sigma_{\rm \nu_e Pb}$ are canceled out in Eqs.~(\ref{eq:Intro:sin22q=}), (\ref{eq:App:hatLambda=}) and $\sin^2 2 \theta_D$ can be measured largely free from systematic uncertainties.
Although it is difficult to know the absolute DAR $\nu_e$ flux with precision better than $10~\%$, the deficit can be measured precisely by comparing the neutrino flux detected in near and far detectors.

Now we assume that we performed an experiment and observed the same number of $\nu_e$ events in both the near and the far detectors,  
\begin{equation}
 n_F = n_N \equiv n_{\rm obs}^D
 \label{eq:App:Nobs=NF=NN}
\end{equation}
In this case, from Eq.~(\ref{eq:Intro:sin22q=}),  the central value of the measured mixing angle is 0 and the result of the measurement is written as follows.
\begin{equation}
 \sin^2 2\theta_D = 0 \pm \delta \sin^22\theta_D
\end{equation}
where $\delta \sin^22\theta_D$ is the error of the $\sin^2 2\theta_D$ measurement. 
Taking only statistical uncertainty into account, that is, $\delta n_F = \delta n_N = \sqrt{n_{\rm obs}^D}$, 
the $1\sigma$ error of $\sin^2 2\theta_D$ is, 
\begin{equation}
 \delta \sin^22\theta_D = 
 \frac{1}{ \left| \hat{\Lambda}_{\rm O}[L_F;\Delta m_D^2] - \hat{\Lambda}_{\rm O} [L_N; \Delta m_D^2] \right|} 
 \sqrt{\frac{2}{n_{\rm obs}^D} }~.
\end{equation}
The expected number of observed 
$\nu_e^\mu$ event is, from Eqs.~(\ref{eq:App:epsilinnuemu}) and (\ref{eq:Nnuemu=}), 
\begin{equation}
 n_{\rm obs}^D=   \epsilon_{\nu_e^\mu}^D  N_{\nu_e^\mu}^0 =  0.1 \times 530,000 =53,000~.
 \label{eq:App:Nobsnuemu=}
\end{equation}
Therefore,
\begin{equation}
 \delta \sin^22\theta_D = 
 \frac{6.1\times 10^{-3}}{ \left| \hat{\Lambda}_{\rm O}[L_F;\Delta m_D^2] - \hat{\Lambda}_{\rm O} [L_N; \Delta m_D^2] \right|} ~.
\end{equation}

Since $\hat{\Lambda}_{\rm O}$ is a function of $\Delta m_D^2$, the upper limit of $\sin^22\theta_D$ depends on $\Delta m_D^2$.
 
\begin{figure}[htbp]
\centering
\includegraphics[width=80mm]{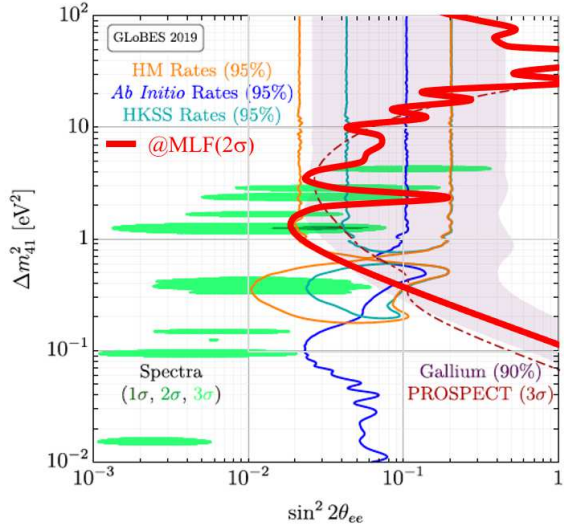}
\caption{\small{
$2\sigma$ statistical upper limit of $\sin^22\theta$ from the disappearance measurement (thick red line) and results of other experiments~\cite{Berryman20}. 
The baseline and the target lead mass of the near and far detectors are (10~m, 11~ton) and (30~m, 100~ton), respectively. 
1MW beam operation($P_B^*$), 5years data taking time ($T$) and 10\% of event selection efficiency ($\epsilon_{\nu_e^\mu}^D$) are assumed. 
}}
\label{fig:Disappearance_sensitivity}
\end{figure}

Fig.~\ref{fig:Disappearance_sensitivity} shows the $2\sigma$ sensitivity of this experiment comparing with results of other experiments. 

\subsubsection{Detection of $\nu_\mu \to \nu_e^\pi$ appearance and a possibility of clean CPV measurement}
$\nu_\mu$ produced in the $\pi^+$ DAR may oscillate to $\nu_e^\pi$ as shown in Fig.~\ref{fig:intro:DaR_nu_Oscillation} and we may be able to detect the $\nu_e^\pi$ by using the lead target.
Since the $\nu_\mu$ is produced in  two body decay of $\pi^+$ at rest, it has a unique energy ($E_0= 30$~MeV) and 
$\nu_e^\pi$ has also the unique energy and its energy spectrum is expressed as the delta function. 
\begin{equation}
 \hat{\Gamma}_{\nu_e^\pi} (E_\nu) = \delta(E_\nu -E_0)~.
\end{equation}

Therefore, the number of $\nu_e^\pi$ generated by the neutrino oscillation is  
 \begin{equation}
  \begin{split}
   N_{\nu_e^\pi}^A [L]&=   \sin^22\theta_A  \times   T \phi_{fA} 
   \int \hat{\Gamma}_{\nu_e^\pi} (E_\nu) \sigma_{\rm \nu_e Pb} (E_\nu)  \sin^2 \Phi_A dE_\nu \\
   &= T \phi_{fA} \sigma_{\rm \nu_e Pb}^0 \sin^22\theta_A\sin^2 \left( \Phi_A^0 [L] \right)
   \end{split}
 \end{equation}
 where $\sigma_{\rm \nu_e Pb}^0 =\sigma_{\rm \nu_e Pb}[E_0]$ and $\Phi_A^0 [L]= \Delta m_A^2 L/4 E_0$. 
 The sub- and super-script  ``$A$,, represents the parameters for appearance.  

Some experiments indicate there is sterile neutrino oscillation with $\sin^22\theta_A \sim O[10^{-3}]$ and 
$\Delta m_A^2 \sim O[1{\rm ~eV^2}]$~\cite{LSND98, MiniBooNE21}.
One possible problem for the small $\nu_e^\pi $ appearance measurement is that $\nu_e^\mu$  may become physics background for $\nu_e^\pi$ detection. 
However,  $\nu_e^\pi$  can statistically be separated from $\nu_e^\mu$  by making use of the differences of production time, energy spectra and event topologies.

Here, we assume to detect only neutrinos that are generated within the first beam pulse shown in Figs.~\ref{fig:Intro:beam_timing}  and \ref{fig:Intro:Timing_cut} to obtain better S/N event sample. 
The efficiencies of the beam pulse timing cut for $\nu_\mu^\pi$ and $\nu_e^\mu$ are 
\begin{equation}
 \begin{split}
 \varepsilon_{\nu_e^\pi}^{bt} [0<t<w] &= 1-\frac{\tau_\pi}{w}\left(1-e^{-(w/\tau_\pi)}\right) = 0.746,  \\
 \varepsilon_{\nu_e^\mu}^{bt} [0<t<w] &= 1-\frac{\tau_\mu^2 \left(1-e^{-(w/\tau_\mu)} \right)-
 \tau_\pi^2 \left( 1-e^{-(w/\tau_\pi )}\right)}{w(\tau_\mu -\tau_\pi)} = 0.0137,
 \end{split}
 \label{eq:App:epsilont}
\end{equation}
respectively, where $t$ is the time from the start of the beam pulse, $w$(=100ns) is the pulse width and $\tau_\pi$ and $\tau_\mu$ are the lifetime of $\pi^+$ and $\mu^+$, respectively. 
Therefore, $\nu_e^\mu$ contamination can be reduced to 1.8\% by the beam timing cut only. 

\begin{figure}[htbp]
\centering
\includegraphics[width=80mm]{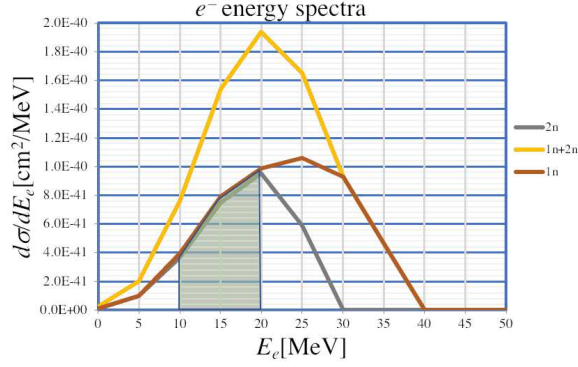}
\caption{\small{ Green area shows energy window (10MeV$\textless E_e\textless$ 20MeV) and requirement of $1n$ emission for the 
$\nu_e^\pi$ selection. 
The efficiency of this cut for the $\nu_e^\mu$ events (yellow line) is $\sim$20\%.
}}
\label{fig:Ecut_for_nuepi}
\end{figure}

\begin{figure}[htbp]
\centering
\includegraphics[width=80mm]{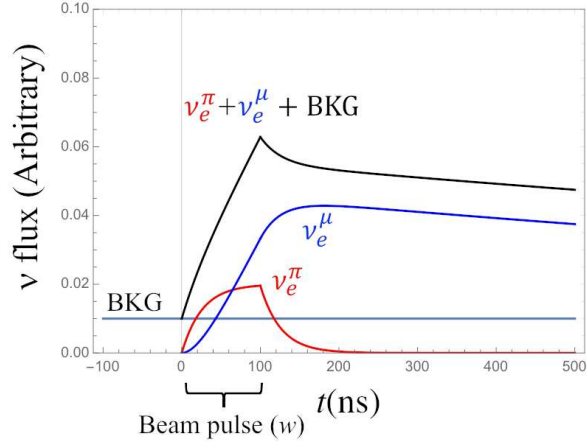}
\caption{\small{ Time structure of $\nu_e^\pi$ and $\nu_e^\mu$ fluxes after the $e^-$ energy and $1n$ cut. 
$\sin^22\theta_A\sin^2\Phi_A \sim 0.005$ is assumed. 
The beam pulse width is 100~ns. 
Appearance of $\nu_e^\pi$ can be identified as the excess around the beam pulse timing. 
Beam uncorrelated backgrounds can be estimated from the event rate at $t<0$~ns .
The contamination of $\nu_e^\mu$ can be estimated precisely from the $\nu_e^\mu$ event rate at $t>200$~ns. 
}}
\label{fig:App:nue_time_structure}
\end{figure}

Moreover, since the energy of $\nu_e^\pi$ is 30~MeV, the energy of the generated $e^-$ is less than 20~MeV because at least 10~MeV is necessary to free one neutron  as shown in Eq.~(\ref{eq:Signal:DM=D208-D207-Dn}).
In addition, as seen in the table of Fig.~\ref{fig:App:nu-Pb_Xsection}, the 90\% of the $\nu_e^\pi$ reaction is $1n$ emission mode. 
By setting the energy window for the $e^-$ detection as 10MeV$\textless E_e \textless$20MeV and choosing only $1n$ emission event\footnote{We assume $1n$ and $2n$ emission events can be separated in large detectors.}, the efficiency for the $\nu_e^\mu$ signal becomes  $\epsilon_{\nu_e^\pi}^{En}\sim 20\%$ as shown in Fig.~\ref{fig:Ecut_for_nuepi}, while the efficiency for $\nu_e^\pi$ is $\epsilon_{\nu_e^\mu}^{En}\sim 80 \%$. 
Therefore, $\nu_e^\mu$ contamination is reduced to  $0.018\times 0.2/0.8\sim 0.5 \%$.
Fig.~\ref{fig:App:nue_time_structure} shows the time structure of $\nu_e^\pi$ and $\nu_e^\mu$ generation after the $e^-$ energy and $1n$ cut. 
$\sin^22\theta_A\sin^2\Phi_A=0.005$ is assumed. 
Appearance of $\nu_e^\pi$ can be identified as the excess around the beam pulse timing. 
Beam uncorrelated backgrounds can be estimated from the event rate at $t<0$~ns .
The contamination of $\nu_e^\mu$ can be estimated precisely from the $\nu_e^\mu$ event rate at $t>200$~ns. 
The oscillation probability measurement becomes insensitive to the absolute scale of the $\nu_e$-Pb cross section 
because it is obtained from the ratio of the number of $\nu_e^\pi$ events and the number of 
$\nu_e^\mu$ events as shown below.

The number of $\nu_e^\pi$  events to detect can be written as 
\begin{equation}
 \begin{split}
  n_{\nu_e^\pi}^A =&   \sin^22\theta_A \sin^2 \Phi_A^0  \\
   & \times \frac{1}{2}  T  \phi_{fA} \sigma_{\rm \nu_e Pb}^0
   \epsilon^* \epsilon_{\nu_e^\pi}^{bt}[0\textless t \textless 100{\rm ns}] 
   \epsilon_{\nu_e^\pi}^{En}[10\textless E_e \textless 20{\rm MeV}:1n]
   \end{split}
   \label{eq:App:nnuepiA}
 \end{equation}
 and and $\nu_e^\mu$ to detect can be written as
 \begin{equation}
   n_{\nu_e^\mu}^A = \frac{1}{2}  T  \phi_{fA} \braket{\sigma_{\rm \nu_e^\mu Pb}}
   \epsilon^* 
   \epsilon_{\nu_e^\mu}^{bt}[0\textless t\textless 100{\rm ns}] 
   \epsilon_{\nu_e^\mu}^{En}[10\textless E_e\textless 20{\rm MeV}:1n], 
    \label{eq:App:nnuemuA}
\end{equation}
respectively, 
where $\epsilon^*$ is the efficiency for cuts other than the timing, $E_e$ and number of neutron. 
Since the average energy of $\nu_e^\mu$  is similar to $E_0$, we assume $\epsilon^*$ is the same for both $\nu_e^\pi$ and $\nu_e^\mu$ detections. 
The overall factor 1/2 comes from the condition that we use only the first beam pulse out of two successive pulses. 

The number of observed $\nu_e$ events is the sum of $\nu_e^\pi$ and $\nu_e^\mu$; 
$n_{\rm obs}^A = n_{\nu_e^\pi}^A + n_{\nu_e^\mu}^A$. 
Therefore,  from Eq.~(\ref{eq:App:nnuepiA}), the oscillation probability can be obtained as 
\begin{equation}
 \sin^2 2\theta_A \sin^2 \Phi_A^0 
 =\frac{(n_{\rm obs}^A - n_{\nu_e^\mu}^A)}
{(T/2) \phi_{fA}\sigma_{\rm \nu_e Pb}^0 
 \epsilon^*
 \epsilon_{\nu_e^\pi}^{bt}[0<t<100{\rm ns}]
 \epsilon_{\nu_e^\pi}^{En}[10<E_e<20{\rm MeV}:1n] }
 \label{eq:App:sin22qAsin2PA0=}
\end{equation}
The denominator can be related to $n_{\nu_e^\mu}^A$ using Eq.~(\ref{eq:App:nnuemuA}) as 
\begin{equation}
 \begin{split}
 {\rm Denominator} &=n_{\nu_e^\mu}^A  \frac{\sigma_{\rm \nu_e Pb}^0}{\braket{\sigma_{\rm \nu_e^\mu Pb}}}
 \frac{ \epsilon_{\nu_e^\pi}^{bt}[0<t<100{\rm ns}] \epsilon_{\nu_e^\pi}^{En}[10<E_e<20{\rm MeV}:1n] }
 {\epsilon_{\nu_e^\mu}^{bt}[0\textless t\textless 100{\rm ns}] 
   \epsilon_{\nu_e^\mu}^{En}[10\textless E_e\textless 20{\rm MeV}:1n]} \\
  & \sim 150 n_{\nu_e^\mu}^A 
   \end{split}
\end{equation}
where, from Eq.~(\ref{eq:App:<sigma>}) and the table in Fig.~\ref{fig:App:nu-Pb_Xsection},
\begin{equation}
  \frac{\sigma_{\rm \nu_e Pb}^0}{\braket{\sigma_{\rm \nu_e^\mu Pb}}} 
  =\frac{28.2\times 10^{-40} [{\rm cm^2}]} {37.5 \times 10^{-40} [{\rm cm^2}]}=0.75
\end{equation}
is used. 
Therefore, Eq.~(\ref{eq:App:sin22qAsin2PA0=}) becomes 
\begin{equation}
\sin^2 2\theta_A \sin^2 \Phi_A^0 \sim 
6.7\times 10^{-3} \left(\frac{n_{\rm obs}^A}{n_{\nu_e^\mu}^A} -1 \right)
\label{eq:sin22qAsin2Pa0=2}
 \end{equation}

Now we assume that we performed the experiment and observed the expected number of $\nu_e^\mu$ background events;
\begin{equation}
 n_{\rm obs}^A = n_{\nu_e^\mu}^A
\end{equation}
In this case,  from Eq.~(\ref{eq:sin22qAsin2Pa0=2}), the central value of the measured mixing angle is 0 and the results of the measurement is written as follows, 
\begin{equation}
 \sin^22\theta_A = \frac{1}{\sin^2\Phi_A^0}\left( 0 \pm \frac{6.7\times 10^{-3}}{ \sqrt{n_{\nu_e^\mu}^A}} \right). 
\end{equation}
where the error is the statistical uncertainty of $n_{\rm obs}^A$. 
$n_{\nu_e^\mu}^A$ can be related to $n_{\nu_e^\mu}^D$ as follows. 
\begin{equation}
 \begin{split}
 n_{\nu_e^\mu}^A &= n_{\nu_e^\mu}^D \times 
 \frac{\epsilon_{\nu_e^\mu}^{bt}[0\textless t\textless 100{\rm ns}] 
   \epsilon_{\nu_e^\mu}^{En}[10\textless E_e\textless 20{\rm MeV}:1n]}
   {2\epsilon_{\nu_e^\mu}^{bt}[1.5\textless t\textless 5.5{\rm \mu s}] 
   \epsilon_{\nu_e^\mu}^{En}[10\textless E_e\textless 40{\rm MeV}:1n+2n]} \\
   &=n_{\nu_e^\mu}^D\times \frac{0.0137 \times 0.2 }{2\times 0.49 \times 0.77}=195
   \end{split}
\end{equation}
Therefore, 
\begin{equation}
  \sin^22\theta_A = \frac{1}{\sin^2\Phi_A^0} \left( 0 \pm 4.8\times 10^{-4} \right)~.
\end{equation}
The result corresponds to the 2$\sigma$ significant upper limit
\begin{equation}
  \sin^22\theta_A < \frac{9.6 \times 10^{-4}}{\sin^2\Phi_A^0}~. 
\end{equation}

The statistical contamination of $\nu_e^\mu$ in the $\nu_e^\pi$ event selection window can precisely be estimated by the off-timing $\nu_e^\mu$ data. 
In addition, since the oscillation probability is measured from the ratio of observed $\nu_e^\pi$ and $\nu_e^\mu$ events, it is  insensitive to the error of the absolute scale of the $\nu_e$-Pb cross section.

\begin{figure}[htbp]
\centering
\includegraphics[width=80mm]{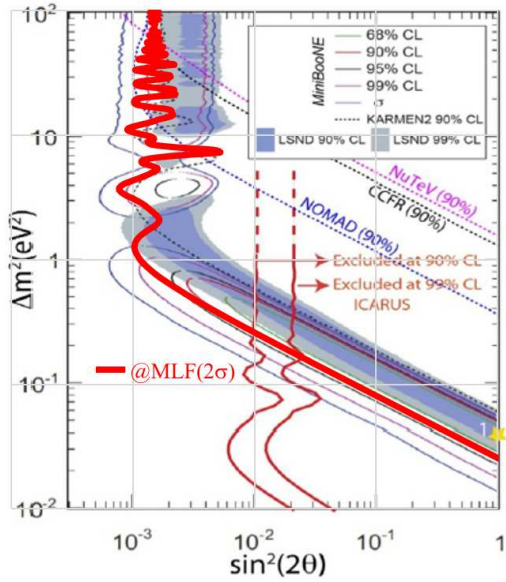}
\caption{\small{
Thick red line shows statistical $2\sigma$ upper limit of $\sin^22\theta$ for the $\nu_\mu^\pi \to \nu_e^\pi$ appearance 
measurement~\cite{JSNS2-TDR17, Antonello13, MiniBooNE21}. 
Assumed detector configuration is, 100~ton lead target far detector at 30~m baseline and 11~ton lead target near detector at 10~m.
1~MW beam power, 100~ns beam pulse width, 5~years data taking are also assumed. 
 A smearing effect due to finite detector size is taken into account. 
}}
\label{fig:Appearance_sensitivity}
\end{figure}

In actual experiment, the oscillation probability will be measured by combining near detector data and far detector data. 
Near detector is sensitive to higher $\Delta m^2$ region and the Far detector is sensitive to lower $\Delta m^2$ region.
By combining the both measurements, the sensitive $\Delta m^2$ region is enlarged.
The $\chi^2$ analysis shows statistical $2\sigma$ upper limit of $\sin^22\theta_A$ is expressed as 
\begin{equation}
 \sin^22\theta_A < \frac{9.6\times 10^{-4}}{\sqrt{\sin^4 \Phi_{AF}^0+\sin^4\Phi_{AN}^0}}
\end{equation}
where $\Phi_{AF}^0 = \Delta m_A^2 L_F/4E_0$ and so on. 
Fig.~\ref{fig:Appearance_sensitivity} shows the sensitivity of the $\nu_e$ appearance measurement.

If both $\overline{\nu}_\mu^\mu \to \overline{\nu}_e^\mu$ and $\nu_\mu^\pi \to \nu_e^\pi$ oscillations are observed, it is possible to test CP violation of the sterile neutrinos. 
Note that the number of $\overline{\nu}_\mu^\mu$ and $\nu_\mu^\pi$ are the same and 
it is not necessary to know absolute neutrino flux to measure the CP asymmetry. 
In addition, the $\nu_\mu^\pi \to \nu_e^\mu$ oscillation probability measurement is insensitive to the absolute scale of $\nu_e$+Pb cross section and 
$\overline{\nu}_\mu^\pi \to \overline{\nu}_e^\mu$ oscillation measurement can be done using well known IBD cross section.
Therefore, clean CPV measurement is possible from the measurements of the DAR $\nu_e$ and 
$\overline{\nu}_e$ measurements. 

\subsection{Detection of Supernova  $\nu_e$}
\begin{figure}[htbp]
\centering
\includegraphics[width=80mm]{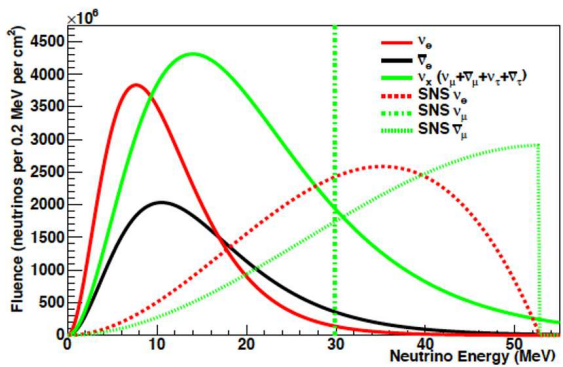}
\caption{\small{ 
Typical expected supernova flux spectra (solid lines) for different flavors, which are integrated up to $\sim$ 15 seconds after burst. 
Dashed lines are DAR neutrino flux spectra integrated for one day at 30 m from the SNS target at Oak Ridge National Laboratory \cite{Bolozdynya1211}.
}}
\label{fig:SN_neutrino}
\end{figure}

Fig.~\ref{fig:SN_neutrino} compares the energy flux spectra of supernova neutrinos, which is the integration of neutrino energy flux over the duration of the burst,  and the DAR neutrinos, . 

The SN flux spectra are taken from a convenient representation of SN neutrinos to be calculated from neutrino transport equation \cite{Tamb13},
\begin{equation}
f_{\alpha} \sim \left(\frac{E}{\braket{E}}\right)^{\alpha} e^{ - (\alpha +1)(E / \braket{E})}~,
\end{equation}
where
\begin{equation}
 \braket{E} = \int_{0}^{\infty} dE E f_{\alpha} (E) / \int_{0}^{\infty} dE f_{\alpha} (E)
\end{equation}
is the averaged neutrino energy. The parameter $\alpha$ can be computed to represent the neutrino spectra as follows
\begin{equation}
 \frac{\braket{E^2}}{\braket{E}} = { {2 + \alpha  } \over {1 + \alpha  }}~.
\end{equation}
The parameter $\alpha$ can be adjusted from the multipole energy from the neutrino transport simulation. 
 
In Fig.~\ref{fig:SN_neutrino}, one can note that both spectra are similar to each other. 
We also expect similar neutrino flux from the project. 
Therefore, this experiment could also be a unique facility for SN neutrino detector and may play an important  role of studying SN explosion phenomena. 
Further it could also serve to test the neutrino mass hierarchy with more sophisticated equipment detector because the SN neutrinos may pass the MSW region in the SN, which causes the rapid transition of neutrino flavor due to the matter effect during the neutrino propagation \cite{Ko20}.

\subsection{Study of  decay at rest $\nu_e$-nucleus interactions}
Study of $\nu$-nucleus interaction is important not only because it is a key reaction for the neutrino-process in the SN explosion, but also it may give great opportunity for studying the neutron skin thickness (NST) of nuclei \cite{Payn19}. The neutrino process is a unique nucleosynthesis in the SN explosion for some p-nuclei, like $^{92}$Nb, $^{98}$Tc, $^{138}$La because those nuclei are blocked by stable nuclei, and consequently cannot be produced by the s- and r-process. But most of the neutrino cross sections necessary for the neutrino process are calculated by theoretical estimation \cite{Haya18}, like QRPA and shell models.

Since the abundances of the nuclei are being investigated by meteorite analysis, we need more reliable neutrino cross section data from experimental side. Since the neutrino beam at JSNS2 is a unique beam available for the SN energy, we expect lots of invaluable data for the neutrino-nucleus cross section. 
This weak interaction cross sections are also useful for understanding the  nuclear responses to the weak interaction.

More interesting study is also possible. It is related to the symmetry energy, which is the energy difference between the symmetric matter composed equal neutrons and protons and the neutron matter. 
By the different pressure of each matter, nuclei may have the NST above the symmetric core. Most of the probes used for measuring the NST exploit electro-magnetic or strong interactions, by which they cannot reach to nuclear matter enveloped by neutrons. But Z-boson in the weak interaction may prove the neutron matter, and consequently we may measure the neutron matter via neutrino scattering and deduce the NST by comparing the charge radius from electron scattering. The present proposal using the lead could provide more reliable data of the NST of the target, by which we could constrain the important parameters in the symmetry energy, like the slope parameter, the incompressibility and so on.

\section{Appendix-3: On-site background measurements}
\label{sec:app3}
\subsection{Introduction and PANDA~4+4 detector}
We have performed on-site background measurement at J-PARC MLF during June-July/2021\footnote{The experiment was performed under a MLF user program (Proposal No. 2020BU9901).}.
The purpose of this background measurement was to know the absolute background rate and to
understand the properties of them at candidate locations in order to design the DaRveX experiment. 
In this appendix, some of the outcomes from the background measurements are described. 

\begin{figure}[htbp]
\centering
 \includegraphics[width=15cm]{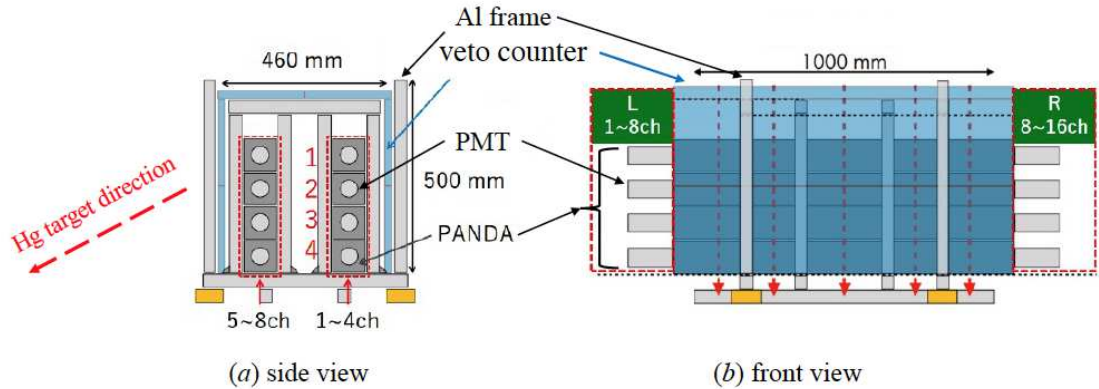}
\caption{\small{Detector structure (PANDA~4+4) for on-site background measurement. 
4+4 PANDA modules are surrounded by six 1~cm thick plastic scintillators.
The Hg target direction is for the case that this detector is placed at the position-A in Fig.~\ref{fig:App:MLF_1st_floor}. }}
\label{fig:App:PANDA4+4}
\end{figure}

Fig.~\ref{fig:App:PANDA4+4} shows the structure of the detector for the background measurements. 
The detector consists of 8 PANDA modules (see Fig.~\ref{fig:Intro:PANDA}) with active mass of 80~kg.  
This is 1/11 of total weight of the DaRveX plastic scintillator. 
4 PANDA modules are piled up to form a tower. 
Two towers stand side by side separated at a distance 20~cm to measure time of flight of particles that hit the both towers. 
The detector is called ``PANDA~4+4,, from its structure. 
The energy calibration of each PANDA module was performed using the penetrating cosmic-ray peak. 
The PANDA~4+4 is surrounded by 6 thin plastic scintillator boards (${\rm 1~cm \times 25~cm \times 150~cm}$)  to veto the cosmic-rays. 
The PANDA modules were read out from the both sides by using 2 inch PMTs (HPK R6410) and the signal shape information was digitized by 500~MHz 8~bit flash ADCs (CAEN V1730D).  
The scintillation light of the cosmic-ray veto-counters were extracted outside using 8 wave length shifting fibers and the signal was read out by MPPC and EASIROC module\cite{Easiroc12}. 

The MLF facility distributes a precise timing signal associated with the beam kicker and current transfer (CT) signal. 
The CT signal indicates the kicked-out beam actually arrived at MLF target.  
Our data acquisition system records those beam timing information and the time between a signal and the kicker can be measured. 

\begin{figure}[htbp]
\centering
 \includegraphics[width=10cm]{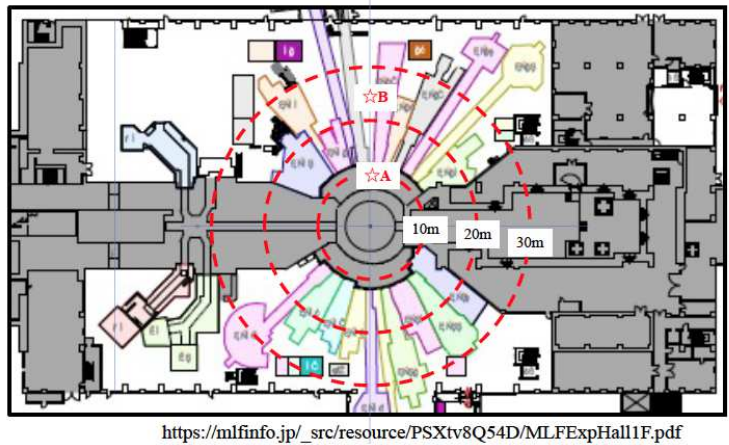}
\caption{\small{The top view of the MLF 1st floor. 
The backgrounds were measured by PANDA~4+4 at the places denoted by ``{\bf A},, and ``{\bf B},, on beam line BL07\cite{Noguchi22}. }}
\label{fig:App:MLF_1st_floor}
\end{figure}

\begin{figure}[htbp]
\centering
 \includegraphics[width=8cm]{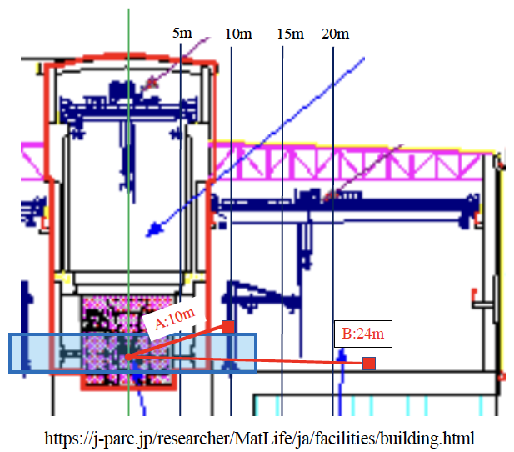}
\caption{\small{The vertical cross section of the MLF beam line viewed from the downstream. 
The position-A is on the front concrete shield with height $\sim$4~m. 
The position-B is on the floor. 
Distance to the mercury target is 10~m and 24~m, respectively. 
 }}
\label{fig:App:BeamLine_SideView}
\end{figure}

We are planning to use the beam line BL07 for DaRveX experiment because there is no existing experiment on this beam line.
The PANDA~4+4 detector measured backgrounds  at the places denoted as ``{\bf A},, and ``{\bf B},, in Fig.~\ref{fig:App:MLF_1st_floor}.
Fig.~\ref{fig:App:BeamLine_SideView} is the vertical cross section of the MLF beam line viewed  from the downstream. 
The position-A is on the front concrete shield with height $\sim$4~m and the position-B is on the floor. 
The distance from the mercury target is 10~m for position-A and 24~m for position-B, respectively.   
Because the expected $\nu_e$+Pb event rate is small, we first seek for possibility to perform the DaRveX experiment at position-A by optimizing the detector and shield structure and then if it turns out necessary, we will consider possibility of the position-B or other places. 

\subsection{Preparatory measurements by CsI counter}
\begin{figure}[htbp]
\centering
 \includegraphics[width=9cm]{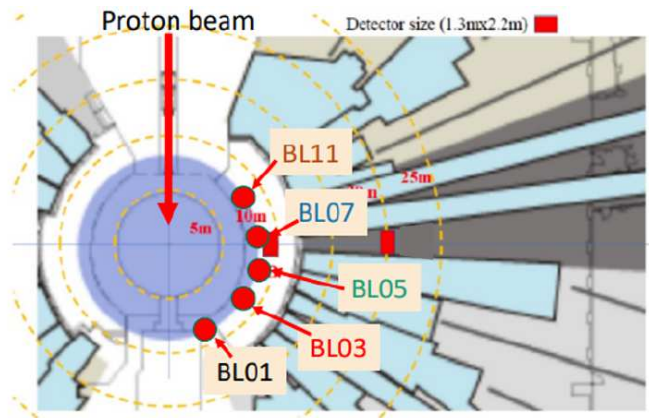}
\caption{\small{The locations of the relative background measurements by CsI counter. 
The location name (BL$n$) corresponds to the neutron BeamLine  {\it n}umber. 
The red squares indicate the footprint size of the DaRveX detector. 
}}
\label{fig:App:Relative_Locations}
\end{figure}

\begin{figure}[htbp]
\centering
 \includegraphics[width=12cm]{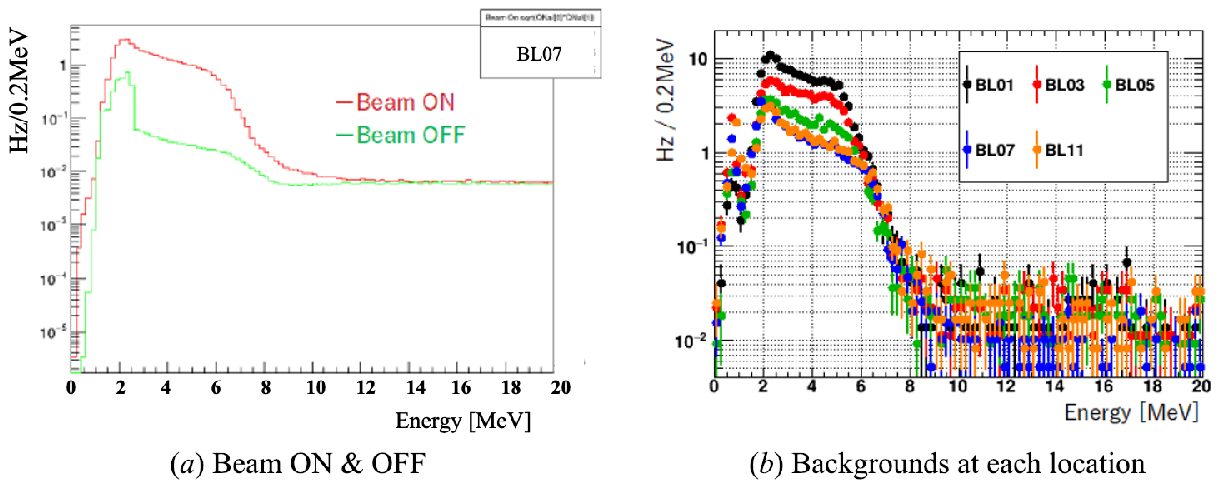}
\caption{\small{$(a)$ Energy spectra measured by the CsI counter for the beam-ON and OFF periods at BL07. 
The peak around 2~MeV in the beam-Off data is $\gamma$-rays from natural U, Th and $^{40}$K decays. 
$(b)$ The energy spectrum of the beam-on CsI data at each location.}}.
\label{fig:App:CsI_data}
\end{figure}

Before the measurements by PANDA~4+4,  we used ${\rm 5~cm \times 6.5~cm \times 28~cm}$ CsI counter to perform preparatory background measurements. 
At first, we measured the relative background rate at locations shown in Fig.~\ref{fig:App:Relative_Locations}. 
Figs.~\ref{fig:App:CsI_data} show the measured data. 
The data were taken with self-trigger and the data contain both on-bunch and off-bunch activities. 
Figs.~\ref{fig:App:CsI_data}$(a)$ shows the energy spectra obtained at location BL07 for beam-ON and OFF periods. 
It clearly shows the beam associated background excess at $E<12$~MeV. 
Fig.~\ref{fig:App:CsI_data}$(b)$ shows the measured energy spectra at each location. 
This measurement shows that the background level at BL07 is relatively smaller than other beam lines.

\subsection{Background measurement with PANDA~4+4 detector}
After the preparatory CsI measurements, we performed background measurements using PANDA~4+4 detector at the positions-A and -B.  
We measured backgrounds at position-A with and without 5~cm thick lead shield on the floor. 
The lead shield was made of 40 lead bricks (5cm$\times$10cm$\times$20cm) that were spread to the area 
(160cm$\times$50cm) beneath the detector and formed 5~cm thick layer. 
On the other hand, we did not use  any neutron shields in this background measurements.
Table~\ref{tab:App:Run_number} summarizes the correspondence between the run number and the data taking conditions.
The background rates measured for prompt and delayed signal windows are also summarized in the 4th and 5th lines in the table. 
\begin{table}[htbp]
\begin{center}
\begin{tabular}{|l||c|c|c|}  
\hline
 ~~~~~~~~~~~~~~~~PANDA~4+4 &~~ Run18 ~~&~~ Run21 ~~&~~ Run33 ~~ \\
  \hline
  \hline
   Position ID & B & A & A \\
 \hline
  Distance from the Hg target [m]   & 24 & 10 & 10  \\ 
\hline 
   Lead Shield (floor; 5~cm thick) & No & No & Yes \\
  \hline 
   \hline   
 Prompt event rate ($10^{-5}$/spill) & $1.1\pm 0.4$ & $6.0 \pm 1.3 $ & $4.0\pm2.3  $\\
 ~ ($20<E<50$MeV, $1.5<dT_B<5.5\mu$s)  & & &\\
     \hline   
 Delayed event rate (/spill) & 0.20 & 4.1 & 2.2 \\
  ~($1.5<E<9$MeV, $10<dT_B<100\mu$s)  & & &\\
      \hline  
 \end{tabular}
\caption{\small{
Correspondence between the run number and the data taking condition,  prompt and delayed background rates for reference energy and timing windows. 
}}
\label{tab:App:Run_number}
\end{center}
\end{table}
%

\subsubsection{Backgrounds for prompted signal}
The neutrino event selection consists of the electron identification as prompt signal and neutron absorption Gd event as delayed signal. 

\begin{figure}[htbp]
\centering
 \includegraphics[width=9cm]{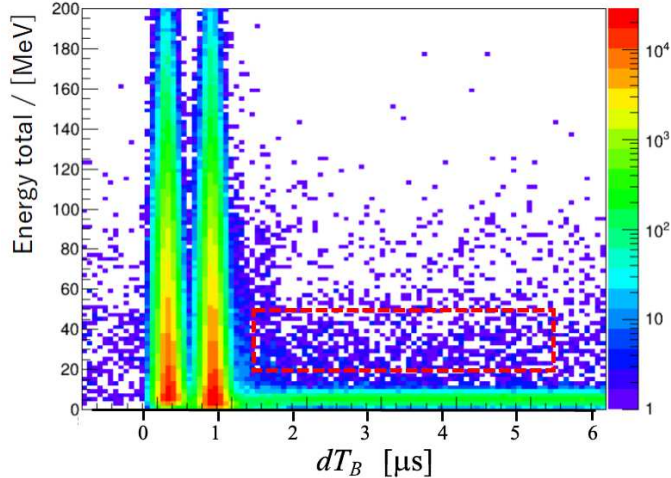}
\caption{\small{The correlation of the timing and the total energy measured by PANDA~4+4 detector (Run33). 
$dT_B$ is the time since the start of the first pulse. 
The two beam pulses and background activities after the beam pulses are clearly seen. 
The $\nu_e$ prompt timing and energy window shown in Table~\ref{tab:App:Run_number} is overlaid as the red-dotted line box.
  }}.
\label{fig:App:nue_window}
\end{figure}

The relation between the  time from the start of the beam pulse ($dT_B$) for the prompt signal region and the total energy of events obtained by the PANDA~4+4  detector at position-A (Run33), is shown in Fig.~\ref{fig:App:nue_window}. 
The two pulse structure and background activities after the beam pulses are clearly seen. 

\begin{figure}[htbp]
\centering
 \includegraphics[width=15cm]{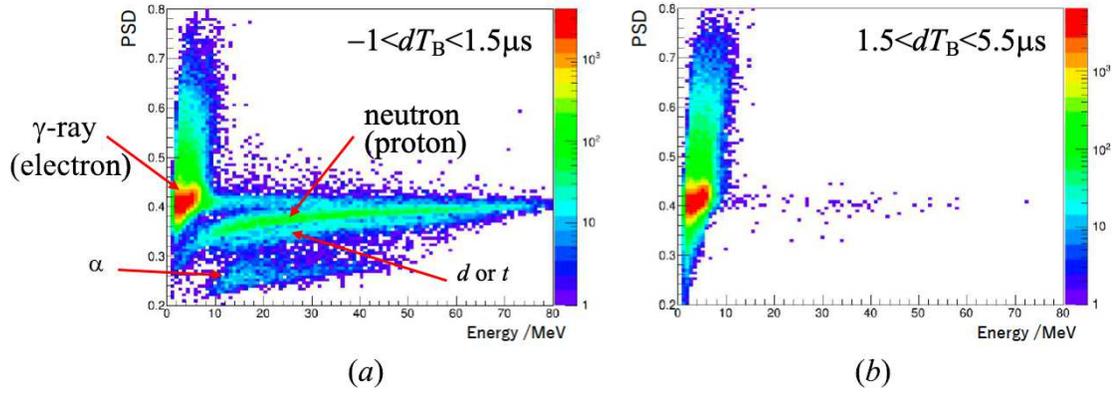}
\caption{\small{ Relation between the PSD parameter and energy, measured by the CsI counter for, $(a)$ on-bunch  and, $(b)$ off-bunch  timings.  
The energetic long island with PSD parameter $\sim 0.35$ in the on-bunch timing is caused by neutron. 
Deuteron (or tritium) and $\alpha$ particle produced are also seen. 
The  neutron signal reduced much in the off-bunch timing. 
The high density blobs at PSD parameter $\sim 0.4$ in both timings are the $\gamma$-ray signal. 
} }.
\label{fig:App:CsI_PSD}
\end{figure}

The CsI counter has capability to distinguish neutron and $\gamma$-ray signals by making use of difference of the pulse shape.  
The neutron signal shape is narrower than $\gamma$-ray signal shape. 
It is called Pulse Shape Discrimination (PSD). 
The time dependent neutron component in the backgrounds were measured at BL07. 
Figs.~\ref{fig:App:CsI_PSD} show the relation between the PSD parameter, that indicates the width of the signal shape,  and energy measured by the CsI counter for,  $(a)$ on-bunch and,  $(b)$ off-bunch timings.  
The Fig.~\ref{fig:App:CsI_PSD}$(a)$ clearly shows the high energy on-bunch background is caused by high energy neutrons. 
The  neutron signal rate reduces much in Fig.~\ref{fig:App:CsI_PSD} $(b)$ but $\gamma$-ray signals exist for both Fig.~\ref{fig:App:CsI_PSD} $(a)$ and $(b)$. 
Neutrons with energy a few MeV or less can enter the timing cut $1.5{\rm \mu s}<dT_B$ and thermalize. 
 The $\gamma$-ray signal in Fig.~\ref{fig:App:CsI_PSD} $(b)$ is expected to be generated by the excited nucleus after absorbing such thermal neutron. 
 
\begin{figure}[htbp]
\centering
 \includegraphics[width=12cm]{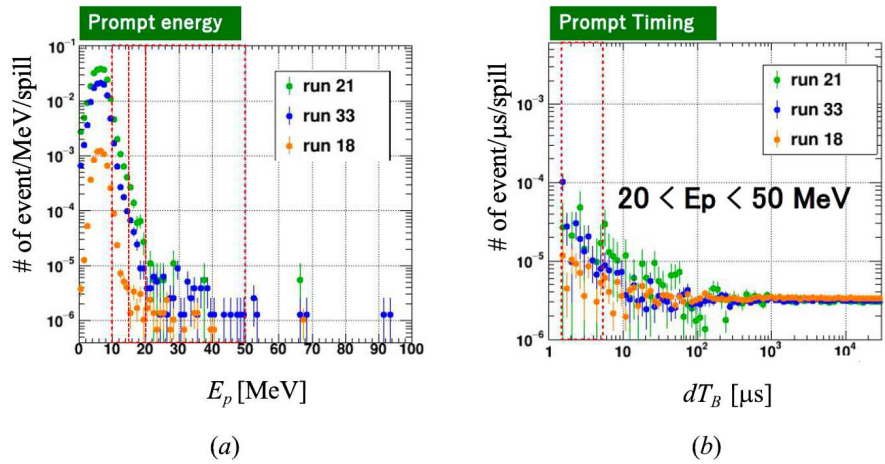}
\caption{\small{ Energy and timing distribution of the prompt signal region. 
$(a)$ Energy spectra of the events whose timing from the start of the beam bunch ($dT_B$) is between 1.5~$\mu$s and 
5.5~$\mu$s.
$(b)$ Timing distributions of the events whose energy ($E_p$) is between 20~MeV and 50~MeV. } }.
\label{fig:App:Prompt_E_T}
\end{figure}

 Fig.~\ref{fig:App:Prompt_E_T}$(a)$  show the energy distribution of the PANDA~4+4 events whose timing is within the prompt signal window  ($1.5{\rm \mu s} < dT_B < 5.5{\rm \mu s}$).  
  In section~\ref{sec:main:event_selection}, it is shown that the electron loses $\sim$5~MeV in the $\nu_e$ target lead sheet in average. 
The background electron is expected also to lose similar amount of energy  in the lead in the DaRveX detector and the 15~MeV$<E_P$ in Table~\ref{tab:Intro:cut_condition} corresponds to 20~MeV$<E_P$ in the energy distribution Fig.~\ref{fig:App:Prompt_E_T}$(a)$. 
Therefore, we define the effective energy window for the prompt event selection is   20~MeV$< E_p<$50~MeV in Fig.~\ref{fig:App:Prompt_E_T}$(a)$. 
Fig.~\ref{fig:App:Prompt_E_T}$(b)$ is the timing distribution of the background signal that has energy within the energy window. 
The 4th line of Table~\ref{tab:App:Run_number} shows the background rate in the prompt event window.
The prompt background rate at baseline $L=$10~m (Run21) is $3.5\sim 8.5$ times more than that of $L=$24~m(Run18), which is consistent with the $L^2$ dependence. 
Although the error is large, the reduction of the background rate due to 5~cm thick floor lead (Run33/Run21) is $0.26 \sim 1.1$. 

For Run33, the prompt background rate is $(1.7\sim 6.3) \times 10^{-5}$/spill or $(37\sim 136)$ event/day. 
Assuming the background rate is proportional to the mass of the plastic scintillator, the background reduction down to 
$(2.4\sim 8.3) \times 10^{-4}$ is necessary to achieve our target signal to noise ratio; S/N=1. 
Therefore, we will need thick neutron and $\gamma$-ray shields and strict event selections requiring event topologies and delayed coincidence. 
%
\subsubsection{Backgrounds for delayed signal}

\begin{figure}[htbp]
\centering
 \includegraphics[width=12cm]{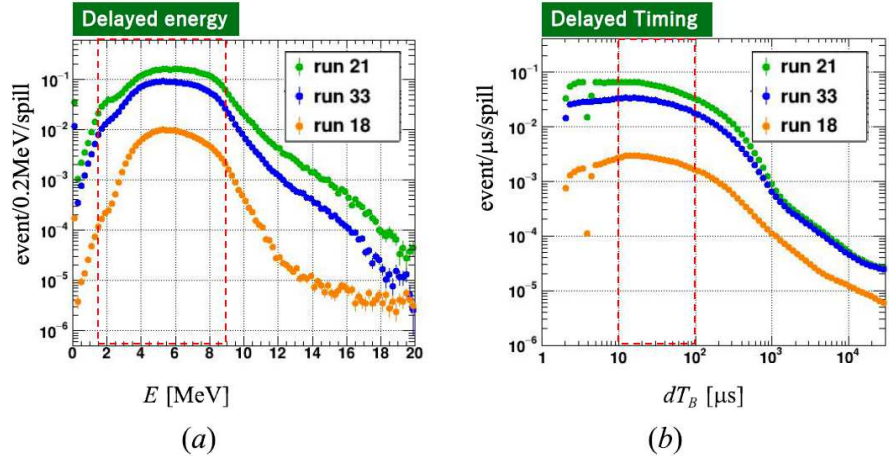}
\caption{\small{ 
The energy and $dT_B$ around the delayed signal. 
 $(a)$ The energy distribution for the events with $10\mu s < dT_B < 100\mu s$.
  $(b)$ The timing distribution for the events of 1.5~MeV$ < E <$ 9~MeV.  
  The delayed event window is indicated by the red-dotted lines.  
  } }
\label{fig:App:BKG_Delayed}
\end{figure}

Figs.~\ref{fig:App:BKG_Delayed} show the energy and timing distribution of the events around the delayed signal window.
These figures show there are significant amount of backgrounds even at $dT_B\sim 100\mu$s. 
The 5th line of Table~\ref{tab:App:Run_number} shows the event rate of the delayed signal. 
The fact that the event rate per spill is greater than one means the delayed coincidence to reduce backgrounds does not work without shields. 
Since the energy is a few MeV and the time scale is much longer than usual lifetime of excited nuclei,  these backgrounds are expected to be $\gamma$-rays from nucleus which absorbed thermal neutron. 

Neutrons with kinetic energy  $E_n < 5$~keV can stay behind at $10\mu{\rm s}<dT_B$. 
Those neutrons are scattered by the nuclei around and thermalize.
Because the Gd-sheets which wrap the PANDA plastic scintillator are exposed to open environment in this background measurement, the thermal neutron can directly reach them and is absorbed by Gd efficiently and emits $\gamma$-rays.
The PANDA scintillator module is thus very sensitive to the thermal neutron background and therefore, neutron shields will be necessary for DaRveX detector. 
In the DaRveX experiment, we will use hermetic-structure Boron sheet layers to block the thermal neutrons. 

\subsection{Summary of the background measurement}
In summary of this on-site backgrounds measurements, we measured the backgrounds for the prompt and delayed signals by using CsI scintillator and PANDA~4+4 plastic scintillator detector at candidate locations in J-PARC MLF. 
The  background rate for the prompt signal timing and energy window showed that it is necessary to reduce background rate down to a few times of $10^{-4}$ by radiation shields and event selections. 
The background rate for the delayed signal window was also measured and it was shown that the thermal neutron shield is important as well as $\gamma$-ray shield. 
We are now planning to measure the effect of thick $\gamma$-ray shields and thermal neutron shields for the real background condition.

\begin{center}
{\large Acknowledgement} 
\end{center}

The on-site background measurement at the MLF of the J-PARC was performed under a user program (Proposal No. 2020BU9901).
DaRveX group thanks the J-PARC staffs for their supports for the on-site background measurements. 
DaRveX group thanks Dr. A.Cabrera (as LiquidO) for providing some electronics modules. 
DaRveX group acknowledges the support of Ministry of Education, Culture, Sports, Science, and Technology (MEXT) and JSPS for grant-in-aid
20H00146.


\end{document}